\documentclass{article}

\usepackage{arxiv}

\usepackage[utf8]{inputenc} 
\usepackage[T1]{fontenc}    
\usepackage[colorlinks,allcolors=blue]{hyperref}
\usepackage{url}            
\usepackage{booktabs}       
\usepackage{amsfonts}       
\usepackage{microtype}      
\usepackage{cleveref}       
\usepackage{graphicx}
\graphicspath{{./figures/}}
\usepackage[comma,authoryear,round]{natbib}
\usepackage{array}
\usepackage{enumitem}
\usepackage{xurl}
\usepackage{authblk}
\usepackage{multicol}
\renewcommand{\undertitle}{}
\renewcommand{\headeright}{}
\renewcommand{\abstractname}{}

\makeatletter
\renewenvironment{abstract}{%
  \quotation
}{\endquotation}
\makeatother

\title{Hardware-Enabled Mechanisms for Verifying Responsible AI Development}

\author[1]{Aidan O'Gara\thanks{Equal contribution}}
\author[2]{Gabriel Kulp\textsuperscript{*} \thanks{Gabriel Kulp is currently serving as a Technology and Security Policy Fellow at RAND; however, the views, opinions, findings, conclusions, and recommendations contained herein are the author’s alone and not those of RAND or its research sponsors, clients, or grantors.}}
\author[3]{Will Hodgkins\textsuperscript{*}}
\author[4]{James Petrie}
\author[5]{Vincent Immler}
\author[6]{Aydin Aysu}
\author[7]{Kanad Basu}
\author[8]{Shivam Bhasin}
\author[9]{Stjepan Picek}
\author[10]{Ankur Srivastava}

\makeatletter
\renewcommand\AB@affilsepx{\quad\protect\Affilfont}
\makeatother

\affil[ ]{
\textsuperscript{1}Oxford University\\
\textsuperscript{2}RAND Corporation, Oregon 
State University\\
\textsuperscript{3}Center for AI Safety\\
\textsuperscript{4}Future of Life Institute\\
\textsuperscript{5}Oregon State University\\
\textsuperscript{6}North Carolina State University\\
\textsuperscript{7}University of Texas at Dallas\\
\textsuperscript{8}Nanyang Technological University\\
\textsuperscript{9}Radboud University\\
\textsuperscript{10}University of Maryland\\
}

\date{}

\begin{document}

\maketitle

\begin{abstract}
\begin{center}\textbf{Abstract}
\end{center}
\vspace{1ex}

Advancements in AI capabilities, driven in large part by scaling up computing resources used for AI training, have created opportunities to address major global challenges but also pose risks of misuse. Hardware-enabled mechanisms (HEMs) can support responsible AI development by enabling verifiable reporting of key properties of AI training activities such as quantity of compute used, training cluster configuration or location, as well as policy enforcement. Such tools can promote transparency and improve security, while addressing privacy and intellectual property concerns. Based on insights from an interdisciplinary workshop, we identify open questions regarding potential implementation approaches, emphasizing the need for further research to ensure robust, scalable solutions.
\end{abstract}

\section*{Executive summary}

Recent years have seen dramatic progress in AI systems' capabilities, in large part achieved through scaling the amount of computation used in the training process. Displaying highly general capabilities, the most sophisticated AI models have the potential to generate significant economic value and help to solve pressing problems facing humanity, but could also facilitate various forms of malicious or harmful use. The potential role of AI hardware in promoting responsible AI development has recently attracted increasing interest from policy-makers and researchers
\citep{Kulp2024, TheBipartisan2024driving}.

This report outlines various hardware-enabled mechanisms that aim to advance certain AI policy goals based on functionality embedded securely into the hardware used to develop and run AI systems. We focus particularly on two classes of mechanisms: mechanisms that can enhance the visibility of governments, external auditors or other relevant stakeholders regarding how AI models are being developed, and mechanisms that can enable enforcement of regulations or agreements relating to high-risk AI development activities. For each mechanism, we discuss the current state of understanding regarding its technical feasibility, potential adversarial attacks, and open research questions. The mechanisms discussed here are not intended to provide an exhaustive list - for more extensive discussion of other options, see \citet{reuel2024open,Scher2024mechanisms} and \citet{petrie2024mechanisms}. This report is based on a workshop held in August 2024 with researchers from hardware security, machine learning, and policy backgrounds to explore these questions.

\textbf{Verifiable AI training and inference.} Hardware-enabled mechanisms could support privacy-preserving and verifiable reports of certain AI workload properties such as quantity of compute used, data used, training techniques or instances of deployments that have occurred. They could also enable policy-makers or external auditors to perform sophisticated evaluations of AI models without direct access to the model weights. These mechanisms could strengthen external oversight and accountability, while respecting the legitimate IP and security concerns of model developers.

\textbf{Verifiable cluster configuration.} High-bandwidth communication between AI accelerators is a key requirement for training state-of-the-art AI systems, making it an important node for verification and enforcement of compliance with safety standards. Understanding which AI accelerators and other network components are connected to perform an AI training run can be valuable in verifying claims made about training computation, techniques or other properties. One potential approach to achieve these goals would be encrypted interconnection between components. This could help to verify cluster configuration and, if warranted by the risks and authorized by relevant regulations, enforce policies by limiting the number of AI accelerators that can be interconnected. In addition to this, such functionality could also enhance the security of AI systems during training and protect against unauthorized access to model weights or training data.

\textbf{Location verification.} In recent years, the US government and allied countries have introduced export controls on AI accelerators to limit their use for military and surveillance purposes by adversaries. Hardware-enabled mechanisms that accurately report a device's location could support well-targeted export controls and reduce the compliance burden by enabling chip owners to provide evidence that their chips have not been diverted to restricted entities. Such mechanisms could facilitate international cooperation to prevent proliferation of advanced AI hardware to untrusted actors such as terrorist groups or rogue states.

\textbf{Offline licensing.} In the future, under scenarios where risks from AI development warrant it or when dealing with less trusted actors, governments and AI developers may choose to enact licensing regimes for AI chips. High-performance AI chips could be designed with cryptographic mechanisms that only allow them to perform operations if they have a license to do so. This kind of licensing system could provide policy-makers and AI developers with more robust tools to enforce safety standards and regulations. In turn, such enforcement capacity would build confidence in mutual compliance, which would enhance incentives for developers and countries to cooperate internationally on AI safety standards.

\textbf{Open challenges.} Effective implementation of hardware-enabled governance mechanisms requires further investigation in several areas. Robust anti-tamper techniques will be essential to secure hardware against both non-intrusive and physical attacks by well-resourced actors, potentially up to and including state-level actors. Mechanism design must also ensure sufficient privacy protections for chip owners, preventing leakage of sensitive IP or unauthorized access to data. Given the heterogeneity of chip designs and of network topologies used by AI developers, mechanisms should also be flexible and adaptable to a variety of cluster configurations. The cost, operational impact and lead time to large-scale development are also important considerations for all these mechanisms.

\tableofcontents
\clearpage

\section{Introduction}
\subsection{Motivation}

AI has the potential to accelerate economic growth and support scientific breakthroughs. However, the current generation of AI systems have highly general capabilities, making AI a dual-use technology. Like any powerful technology, AI carries risks, including the possibilities of misuse by rogue actors and weaponization by nation states, as well as the potential for loss of control over AI systems. These challenges underscore the urgent need for robust safeguards and oversight mechanisms around AI.

\textbf{Misuse of AI capabilities could threaten public safety by lowering the barriers to launching powerful cyberattacks, developing biological weapons, and more}. Researchers have begun to characterize the nascent offensive cyber capabilities of LLMs
\citep{zhang2024cybench,shao2024empirical}. In October 2024, OpenAI found that their newest o1 model ``can help experts with the operational planning of reproducing a known biological threat, which meets our medium risk threshold'' \citep{OpenAI2024OpenAI}. As these dangerous capabilities continue to advance, minimizing risks of misuse by malicious actors is becoming a priority for AI developers and governments.

\textbf{Nation states could also weaponize AI for military purposes in ways that inadvertently increase the risk of armed conflict.} In 2017, Russian President Vladimir Putin said that whichever nation leads in AI ``will be the ruler of the world.'' China's \citeyear{DevelopmentPlan2017} AI Development Plan set the ambitious goal for China to lead the world in artificial intelligence by 2030. The United States is actively pursuing the integration of AI into military systems, investing heavily in AI-enabled autonomous drones \citep{investing2024}. States might eventually become more involved in directly developing AI systems. Just as states have sought to limit the development and use of biological, chemical, and nuclear weapons, they might be interested in avoiding the worst-case outcomes of AI weaponization. Developing hardware-enabled mechanisms to help verify compliance with international agreements could enable cooperation on defining and enforcing acceptable use of military AI systems.

Misuse, weaponization and the disruption of strategic stability are the clearest and most immediate risks to national and international security, but other challenges loom on the horizon. There is the potential threat of loss of control over AI systems, either by voluntarily delegating decision making authority to AI systems to remain economically and militarily competitive \citep{hendrycks2023natural}, or by accidentally developing rogue agents that deliberately seek power in ways unintended by their developers \citep{carlsmith2022power}. AI could also concentrate power in the hands of a small number of people, such as strengthening autocratic regimes by enabling widespread surveillance
\citep{kalluri2023surveillance}, or causing mass automation that leaves individuals dependent on corporations and governments
\citep{acemoglu2019wrong, korinek2024scenarios}. These challenges highlight the need for an approach to AI development that is not solely driven by commercial or competitive pressures, that provides for appropriate external oversight and scrutiny, and that promotes international cooperation on safety standards.

\textbf{Recent advances in AI have been driven to a large extent by growing hardware inputs.} The last decade of AI development has been governed by a simple paradigm: more computing power and more data allow neural networks to acquire more capabilities.
\cite{kaplan2020scalinglawsneurallanguage}
formalized this insight, observing a mathematical power law relationship between the inputs to a neural network training run (compute, data, and parameters) and the network's performance, termed a ``scaling law.'' To reap the benefits of these scaling laws, the amount of computational power used in the largest AI training runs has increased more than $4\times$ annually over the last decade \citep{epoch2023aitrends}. As an example, GPT-4 was trained with roughly $1000,000\times$ more compute than the original GPT model released in 2018, and each generation of GPT models has used roughly $100\times$ more compute than the last \citep{epoch2023aitrends}. The largest data centers currently under construction will cost billions of dollars and house hundreds of thousands of state-of-the-art GPUs
\citep[][Epoch, forthcoming]{NvidiaBroadcom2024Frontier}. The centrality of computing hardware in AI development makes it a promising node for governance.

\begin{figure}[htb]\centering
\includegraphics[width=\textwidth]{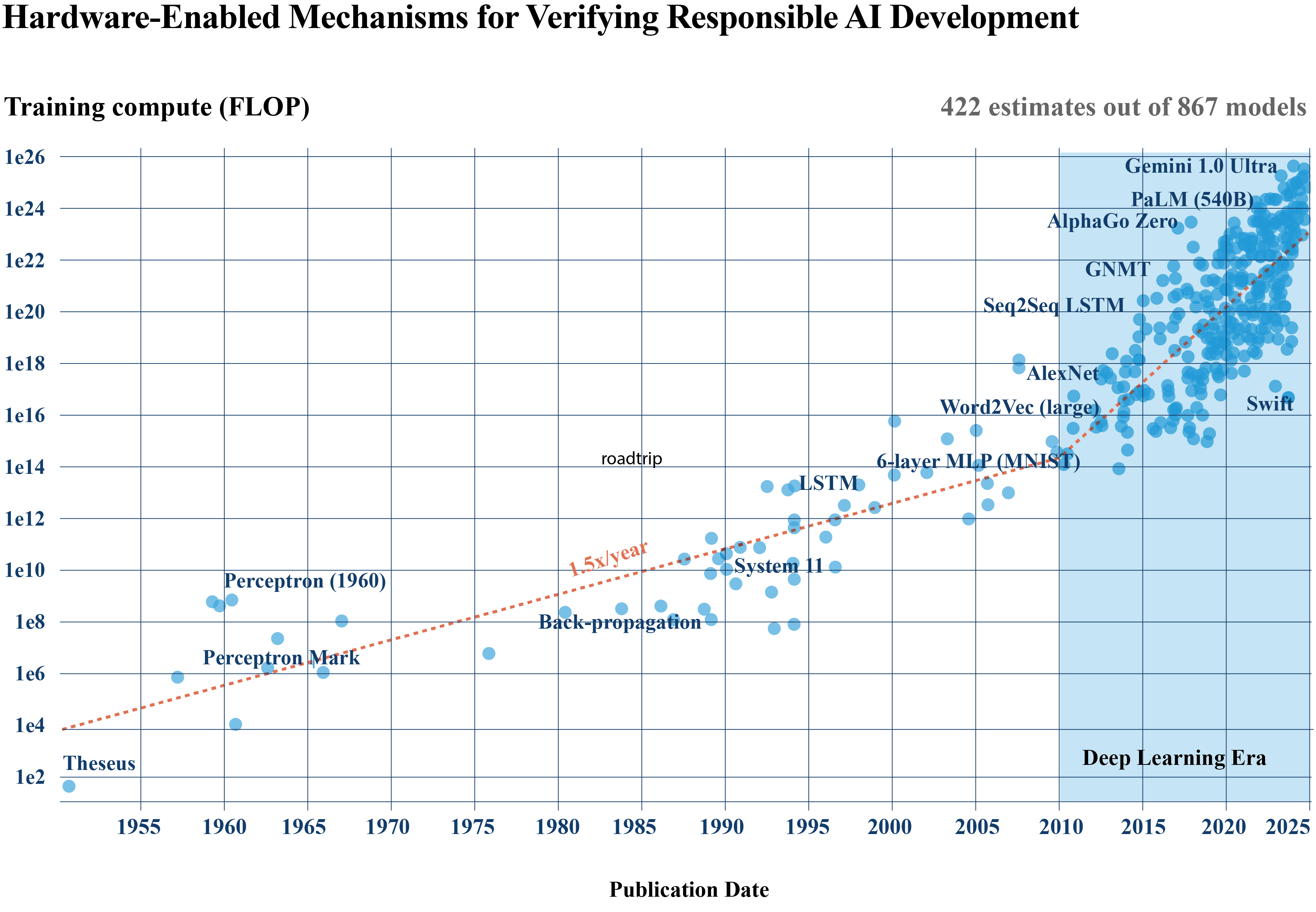}
\caption{Notable AI Models. Adapted from:  \citep{EpochII2024Data}.}
\label{f1.png}
\end{figure}

\textbf{Hardware is a promising lever for AI governance. }AI hardware appears to be a promising node for verifying compliance with responsible AI development practices. While other key inputs to AI development such as data and algorithms are non-physical resources that can be easily duplicated and shared around the world, hardware is a physical and excludable resource. The hardware supply chain is highly concentrated in a few companies---TSMC for chip fabrication, Nvidia for chip design, ASML and others for semiconductor manufacturing equipment---and these companies are predominantly located in countries allied with the United States. As of 2023, high-end data center AI chips represented less than 0.00025\% of all chips produced
\citep{Heim2024What}. Therefore, regulations that target these chips can likely be tailored to have minimal impacts on activities beyond AI development.

\textbf{Hardware-enabled mechanisms should target frontier AI development}. Available evidence from evaluations of current models suggests that they have limited capability to enable malicious actors to cause large-scale harm via cyber-attacks, bioweapons development etc., but that their capabilities in these domains are growing over time. Therefore, frontier AI systems that raise the ceiling of capabilities in such domains are of greatest concern from a public safety and national security perspective. By frontier AI, we refer to general-purpose AI systems at or near the frontier of capabilities, such as Google DeepMind's Gemini 1.5 Pro, Anthropic's Claude 3.5 Sonnet, and OpenAI's o1. Only a handful of actors operate at the frontier of AI development, and each of them spends hundreds of millions of dollars or more in their efforts. They typically use a large number of advanced chips to train such systems (e.g. $>100,000$ Nvidia H100s). It is both more feasible and more desirable to focus governance efforts on these large-scale AI developers relative to the large number of smaller actors that also use AI hardware, such as startups and academic researchers.

\begin{figure}[htb]\centering
\includegraphics[width=\textwidth]{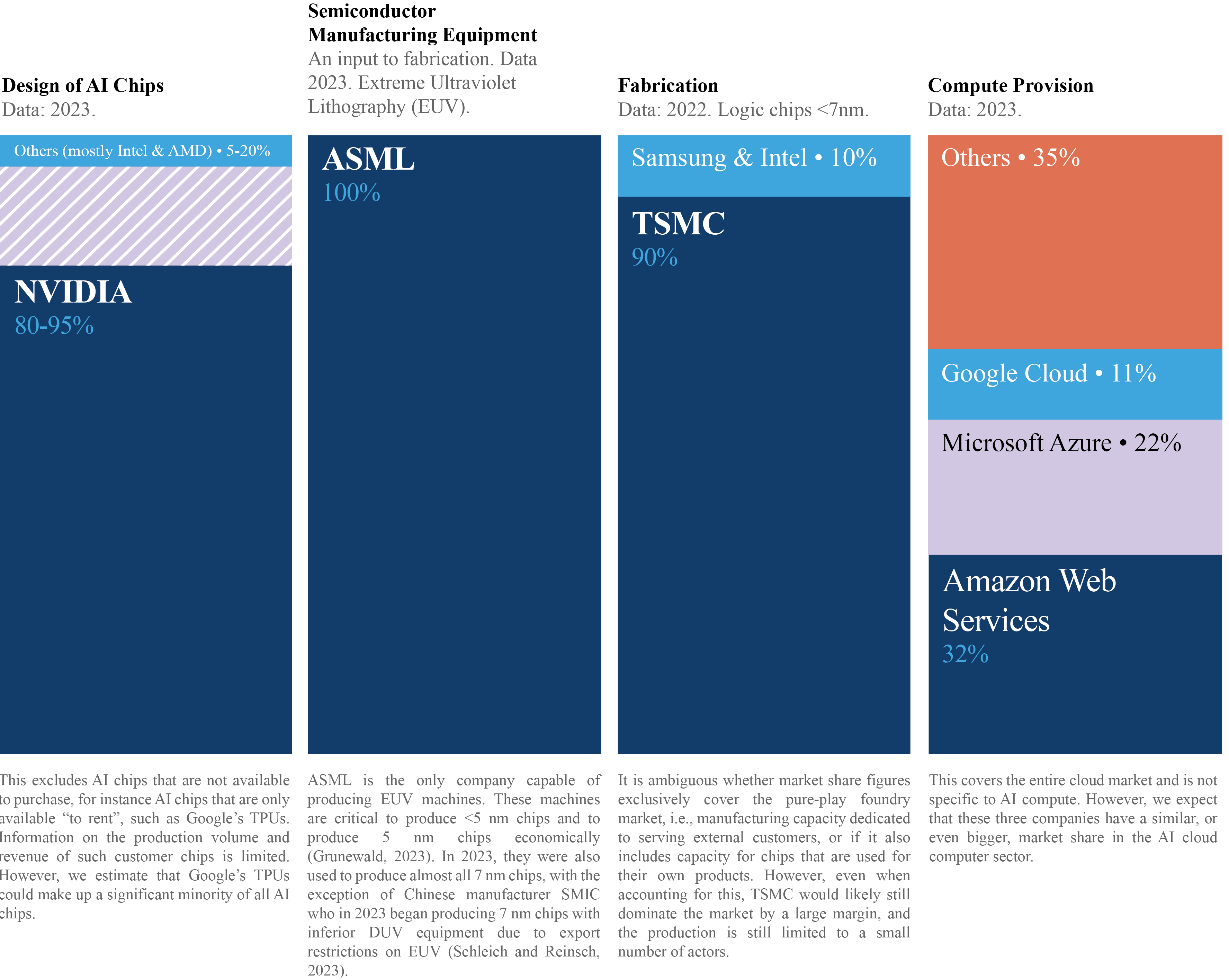}
\caption{Concentration of the AI Chip Supply Chain. Expressed as percentage of total market share.
Adapted from: \citep{sastry2024computing}.}
\label{f2.png}
\end{figure}

\textbf{Hardware-enabled mechanisms could further current and future goals in AI policy.} AI chips could be designed with technical mechanisms that enable AI governance. For example, the United States enforces export controls on high performance AI chips, but there is evidence of large-scale diversion of AI chips to controlled countries \citep{Fist2023Preventing}. Location verification mechanisms on AI chips could allow chip owners to demonstrate in a low-cost manner that they are still in possession of chips and are compliant with export controls. Another policy goal that could be supported by hardware-enabled mechanisms would be reporting of AI development activities to governments or other external bodies. For example, US Executive Order 14110 required developers to report to the federal government any AI system trained using more than 10\textsuperscript{26} operations. Section 3 explains several such technical mechanisms and their purposes.

\textbf{Hardware-enabled mechanisms can create optionality for policy-makers in addressing future governance challenges.} Technical researchers today cannot anticipate all of the potential policy applications of hardware-enabled mechanisms, nor determine whether these policy applications will be desirable in the future. The risks presented by AI systems are evolving over time as capabilities continue to improve. However, given the significant lead time for making potential adjustments to the hardware or software used in AI development and getting these into commercial production, technical research on hardware-enabled mechanisms cannot be purely reactive. We believe the most productive role for technical research today is in creating a set of tools that AI developers and governments can choose to use (or not use) in the future. This increases the value of mechanisms that provide flexibility and can be tailored appropriately based on the risk landscape in coming years.

\textbf{Hardware-enabled mechanisms could reduce the need for expansive export controls.} Firstly, if regulators are sufficiently confident that they can monitor what AI chips are being used for, there may be less of a need to prevent them from ending up in countries perceived as adversaries. This could be an incentive for companies to adopt the measures, as they might then be able to have a larger pool of customers to sell chips to. In practice, the confidence level might need to be extremely high to roll back export controls, and governments are likely to still want multiple layers of defense. However, it is feasible that hardware governance could at least reduce the likelihood of export controls being extended further to additional countries, and enable users in countries that are neither adversaries nor close allies to continue to use AI chips. For instance, if the US government were concerned about chips being smuggled to China through third countries, it might seek to extend controls to those third countries. Reliable location attestation could offer a way of knowing where those chips have ended up and deactivating them in export-controlled countries, allowing companies to continue selling chips to third countries.

\textbf{Hardware-enabled mechanisms could allow companies to enforce their own terms of service.} It might be useful for chip design companies, as well as regulators, to better understand what their hardware is being used for. Nvidia, for instance, makes different types of chips, some of which are intended for large-scale AI training in data centers, and other consumer-type GPUs designed for gaming. There have been examples of customers buying large numbers of the latter type and connecting them to do AI training, in breach of Nvidia’s terms of service. This approach may be cheaper than purchasing GPUs intended for data centers, or enable companies to avoid using products subject to export controls. Companies may therefore stand to gain from monitoring activities, if it enables them to segment the market and restrict products to their intended uses.

\subsection{Objectives}

This report assumes that for large-scale AI computing to be safely integrated into society, several key sub-goals must be met. First, it must be governable, meaning that it should be possible to control access to large-scale AI computing, based on adherence to a shared set of commonsense rules, agreed by democratic nations, to prevent misuse that would create unacceptable risks to public safety and national security.

This level of oversight is desirable because government bodies are more likely to be able to pursue safety without being influenced by commercial incentives. Since the majority of AI development is currently happening in private companies, it is primarily driven by commercial incentives. Given strong competitive pressures to release products quickly in order to keep up with rivals, such incentives may lead companies to release AI models without adequately assessing and mitigating risks. Government bodies may be better positioned to push for strong safety standards, as well as to mediate cooperation between various stakeholders. International coordination is important to ensure rules are consistent across nations, thus avoiding regulatory arbitrage by companies.

A second sub-goal is that the use of large-scale computing must be verifiable; it should be possible to make reliable, externally verifiable claims regarding how computing resources are allocated and used, such as for specific training runs.

This verifiability is essential for the effective implementation of any regulations; it will be difficult to ensure that operators are adhering to rules without techniques and measures that can be trusted to report accurate information about how computing infrastructure is being used. Moreover, verifiable reassurance that actors are abiding by the regulations will be key to preventing security dilemmas. The latter can arise when organizations suspect that their competitors are breaching rules, creating a concern that they may fall behind if they stick to them. This can lead to a scenario in which all parties race ahead, lest rivals gain an advantage, collectively creating a riskier environment for everyone. Verifiable assurances that other parties are adhering to rules can foster a sense of trust and counteract pressures to race.

The third sub-goal for ensuring the safe deployment of large-scale AI is that computing infrastructure must be secure enough to withstand sophisticated attacks, safeguarding critical assets from malicious actors.

The potential power of large-scale AI models might create a strong incentive for some actors to evade hardware-enabled mechanisms and use them for disallowed activities. Mechanisms must therefore be robustly secured against skilled, well-resourced attackers. This will become increasingly important as AI is more widely deployed by a variety of companies and public services. If AI is integrated within critical infrastructure or running essential services, then successful attacks could cause significant harm and disruption.

Lastly, privacy-preserving measures are essential to uphold established data and code privacy norms and minimize the risk of harmful forms of surveillance. If not implemented carefully, hardware-enabled mechanisms could offer governments a way to increase surveillance, infringing on citizens' privacy and potentially suppressing legitimate activities. Companies may also be concerned about maintaining confidentiality of trade secrets. Guardrails that govern how third parties may access information and act upon it will be necessary to address these concerns.

\section{Mechanisms}
\subsection{Overview and scope}

Previous work has underscored the potential for AI hardware to play a critical role in governance by enhancing regulatory visibility, enforcing safety standards, and mitigating risks from irresponsible or malicious AI development \citep{sastry2024computing}. This section explores hardware-enabled mechanisms that can support compliance with regulation, safety standards and corporate policies, strengthen external oversight, and protect intellectual property against potential threats. These mechanisms are particularly relevant for promoting responsible AI development and fostering international cooperation on safety standards.

Hardware-enabled mechanisms refer to solutions that aim to advance certain AI policy goals (e.g. transparency or non-proliferation of hazardous AI systems) based on functionality embedded into the hardware used to develop and run AI systems. In this context, hardware is not restricted to AI accelerators and may include other relevant components such as network interface cards or network switches. Solutions that are embedded in relevant hardware do not necessarily require modifications to existing hardware designs. They may instead be achievable purely through software updates such as changes to firmware or drivers, taking advantage of existing features such as Secure Boot or Trusted Execution Environments (TEEs).

We provide an overview of four key mechanisms for AI governance:
\begin{enumerate}
\item  \textbf{Verifiable AI training and inference}. Privacy-preserving, hardware-enabled mechanisms could enable the generation of trustworthy, verifiable measurements of AI workload properties, such as workload size or operational hours. In addition, these mechanisms could allow policy-makers or auditors to conduct a wide range of evaluations without direct access to model weights, bolstering accountability while safeguarding IP and security concerns.

\item  \textbf{Verifiable cluster configuration.} Understanding the number of AI accelerators connected in high-bandwidth configurations can help to verify claims made about the number of operations or other properties of AI training runs. If warranted by the risks and authorized by relevant regulation, limits on the size of a training cluster could be used to limit high-risk AI training activities while maintaining functionality for smaller-scale training and inference.

\item  \textbf{Location verification.} Secure hardware-based location reporting could support export controls and prevent AI chips from being diverted to untrusted entities. Over time, such mechanisms could underpin international agreements restricting access to advanced AI hardware to actors that adhere to responsible AI development standards. This could simply provide visibility to regulators, or could be combined with other mechanisms to enable geographically targeted policy enforcement, such as geolocking of AI accelerators in prohibited locations.

\item \textbf{Offline licensing.} If warranted by the risk level posed by AI training or inference activities, or when dealing with untrusted actors, chips could be designed to require a license for use. These licenses would be cryptographic keys issued by a regulatory authority that expire after a specified amount of computational work, requiring the chip owner to acquire a new license. These licenses could ensure compliance with safety standards and enable more robust governance in certain scenarios.
\end{enumerate}

While this is not an exhaustive list, these mechanisms represent promising tools to enhance the governance of advanced AI systems and create incentives for international collaboration on AI safety.

\subsection{Verifiable AI training and inference}

AI chips could be designed to generate verifiable digital certificates that attest to key aspects of device operation, such as the size of an AI workload or the number of hours the device has been active. These certificates would support regulators in identifying and monitoring high-risk AI training activities. Certification could be particularly valuable for verifying compliance in lower-trust environments, such as when commercial or geopolitical competitors agree to comply with certain AI standards. Properties that might be verified can be split into at least three sets with distinct technical requirements:
\begin{itemize}
\item  \textbf{Compute accounting} could be facilitated by providing signed measurements of the computational activity carried out as part of an AI training run. Such measurements would be valuable for regulatory compliance with policies that require reporting if a model's training goes above a certain computational threshold (e.g., US Executive Order 14110, EU AI Act). They could also be useful for environmental impact tracking.

\item  \textbf{Workload classification} aims to distinguish AI from non-AI workloads, or AI workloads at different stages of the model lifecycle (design, pre-training, post-training enhancements, and deployment), to ensure that monitoring activities are appropriately targeted.

\item  \textbf{Detailed workload verification} involves verifying properties of the AI model or training dataset, for example verifying whether the dataset contains high-risk content (e.g. data relevant to building weapons) or confirming that certain safety evaluations have been conducted.
\end{itemize}

\begin{figure}[htb]\centering
\includegraphics[width=\textwidth]{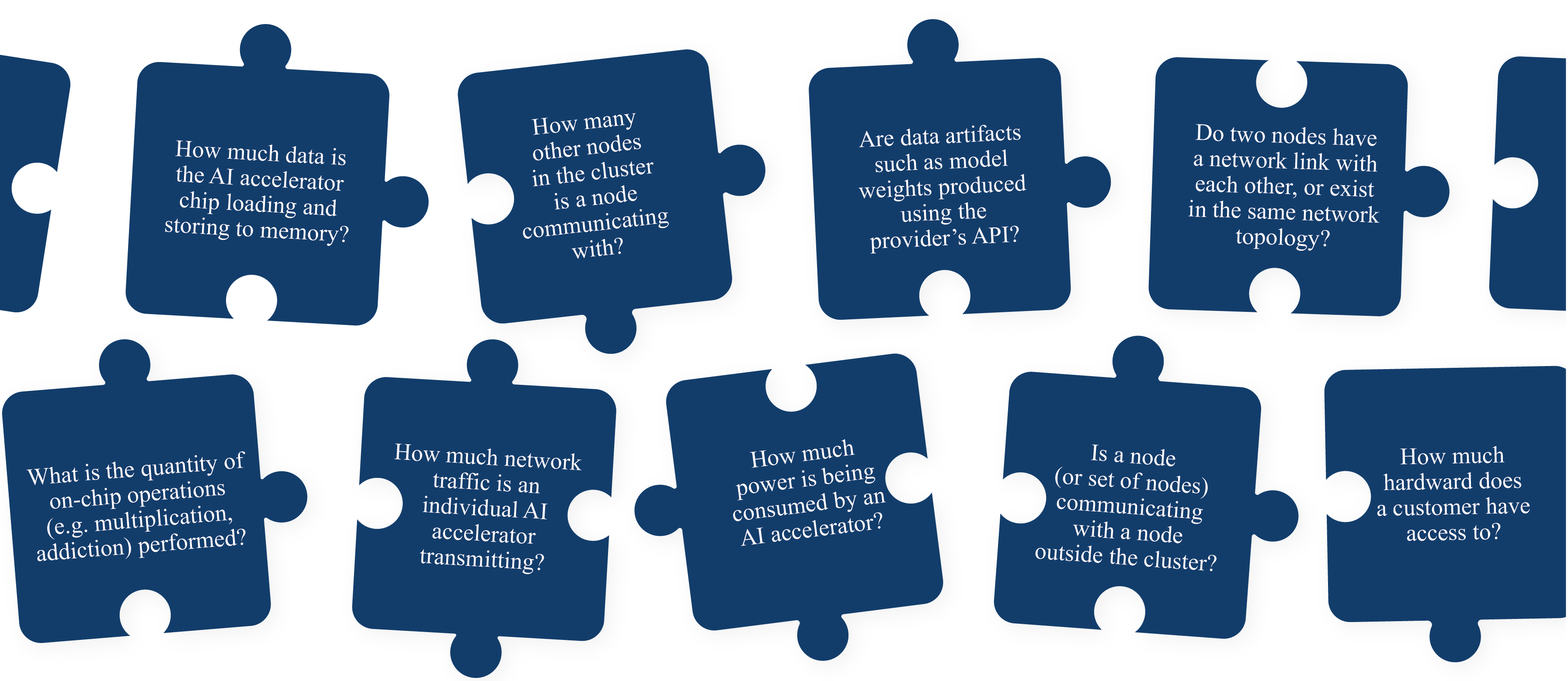}
\caption{Adapted from: \citep{heim2024governing}.}
\label{f3.png}
\end{figure}

\subsubsection{Related work}

\cite{heim2024governing} provide an extensive discussion of methods for workload classification and measurement in AI systems. Some approaches to measuring AI training activities include the use of remote attestation within a secure hardware module, such as a trusted execution environment (TEE), as suggested by
\cite{aarne2024secure}, or the concept of ``training transcripts,'' which involve saving snapshots of model weights during training, as proposed by \cite{shavit2023does}.

\subsubsection{Technical details}

\textbf{Compute Accounting.} To implement compute accounting, the chip would require meters to monitor relevant metrics, such as clock cycles or other indicators of computational activity. These measurements would ideally be collected using a trusted execution environment (TEE) or other secure enclave. The chip would generate a device-specific public and private key pair, ideally through a dedicated security module. The public key would be shared externally before measurements are taken, while the private key would be used to create cryptographic signatures of the device's measurements. This allows a verifier to validate the authenticity of the measurements, confirming that they genuinely originated from the device.

\textbf{Workload Classification.} For workload classification, robust indicators must be identified to distinguish AI-related workloads from other operations. Indicators like the scale and configuration of hardware provide valuable clues to this. For more granular classification, heuristics based on technical characteristics could be manually defined, or machine learning classifiers can be trained using cluster- and node-level data. Frontier-scale AI training, for example, could be identified by unique patterns, such as the number of accelerators, steady peak throughput utilization, and specific communication patterns within nodes that correspond to different forms of parallelization.

Classification could be performed using measurements from individual AI chips or other devices such as network switches. Hardware-backed classification approaches could provide greater confidence that reports genuinely originated from the AI hardware that is intended to be monitored.

\begin{figure}[htb]\centering
\includegraphics[scale=0.12]{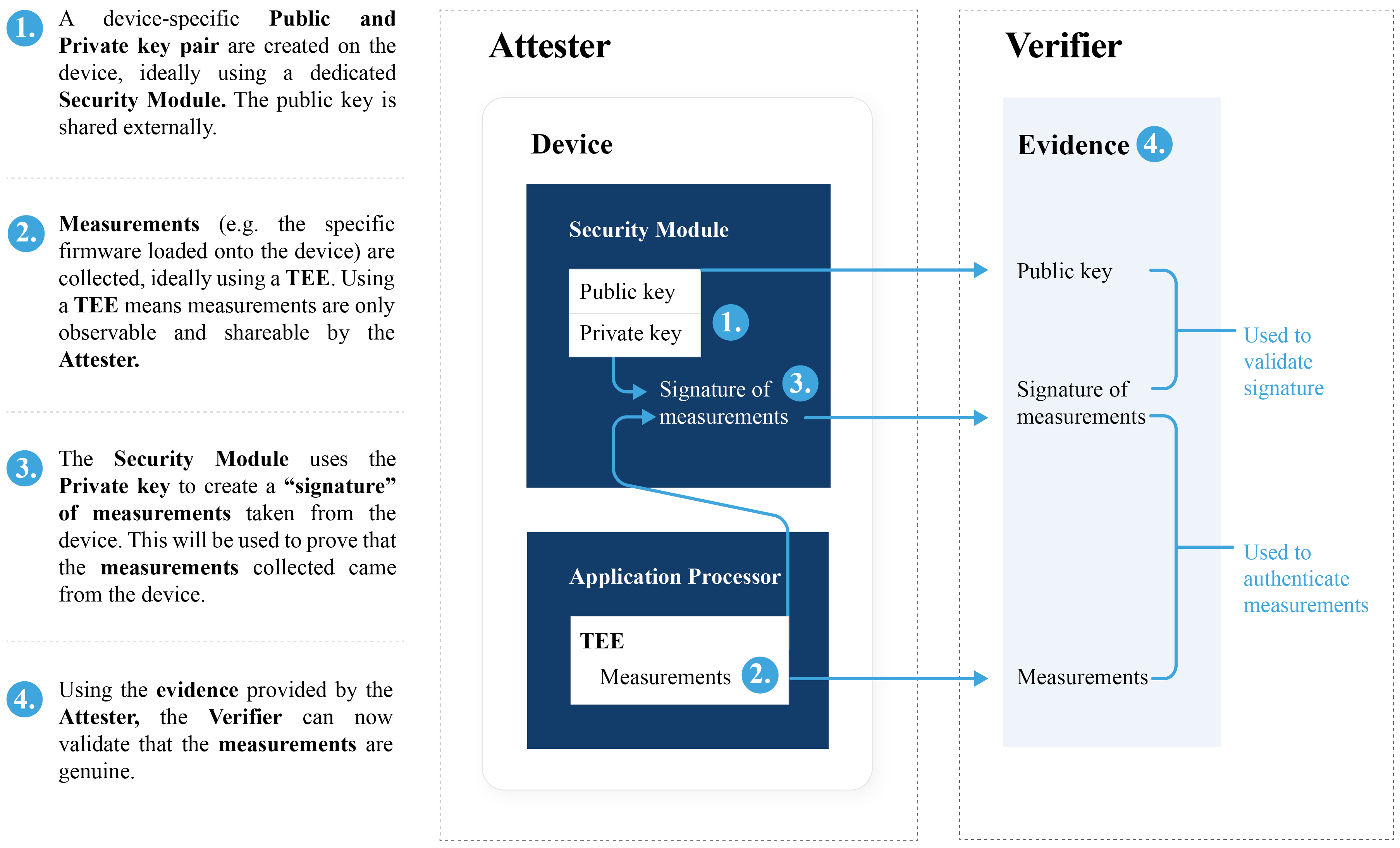}
\caption{Adapted from: \cite{aarne2024secure}.}
\label{f4.png}
\end{figure}

\textbf{Detailed Workload Verification.} Detailed workload verification seeks to provide assurances about governance-relevant properties of a workload without requiring direct access to customer code or data, preserving confidentiality. Techniques like confidential computing can allow customers to prove specific aspects of their workloads to compute providers or regulators.
For instance, customers could demonstrate that they used or avoided particular datasets or performed specific model evaluations. Secure enclaves such as TEEs could help with running these verifications without exposing models or sensitive data. 

\subsubsection{Attacks}

Potential attacks on these monitoring approaches include tampering with secure enclaves, meters, or other components to corrupt or falsify measurements. Adversaries might target the security mechanisms of trusted execution environments (TEEs) or meters to forge or manipulate data.

Adversaries could intentionally modify the computational patterns of their workloads to avoid triggering classification as AI-related tasks. For example, they might introduce noise by adding irrelevant computational activities that make classification more challenging.

They might also fragment their activities, distributing workloads across multiple data centers or compute providers, making it difficult for verifiers to get a clear picture of the total computational effort involved.

\subsubsection{Open research questions}
\begin{itemize}
\item What are the most effective metrics and indicators for accurately classifying AI workloads and performing compute accounting, while minimizing performance overhead?
\item  How can adversarial robustness be improved for these metrics, ensuring resilience against intentional manipulation and evolving AI training techniques?
\item  What are the technical and practical challenges of implementing large-scale cryptographic proofs and secure enclaves (e.g., TEEs) for AI verification across multi-node and multi-GPU systems? What modifications to current hardware, protocols, and security features would be needed to enable widespread use of TEEs for verifying governance-relevant properties of AI training at scale?
\item  How can dataset verification be reliably performed in distributed systems, accounting for challenges like pipeline parallelism and data parallelism, where only subsets of GPUs interact with input data?
\end{itemize}

\subsection{Verifiable cluster configuration}

High-bandwidth interconnection is essential to enable training of frontier AI systems based on current training techniques. In order to ensure that reported measurements of AI training activities are complete, it is important to understand how an AI training cluster was configured and which devices were included. Failing this, adversaries may engage in selective reporting of their training activities and conceal certain aspects. Mechanisms that provide regulators or auditors with improved visibility into the number of AI accelerators used in a particular training run, or that allow AI developers to make verifiable claims about this, would be valuable to support external oversight of compliance with safety standards.

In certain scenarios where the risk of harm is sufficiently high, it may also be desirable to prevent unauthorized actors from aggregating high-performance AI chips into large-scale supercomputers for AI training. To achieve this, restrictions could be implemented on the ability of these chips to interconnect to form clusters. Two potential approaches could achieve this.
The first involves configuring a fixed set of AI chips into a predefined pod that allows high-bandwidth interconnection within the pod, but severely limits external communication with devices outside the pod, making it difficult to combine multiple pods for large-scale AI training. The second approach would loosen the restriction of a predefined pod and instead rely on an adjustable cap on the total number of interconnected AI chips or total interconnect bandwidth between them. This cap could be adjusted depending on which activities a particular chip owner or operator is authorized to conduct.

\begin{figure}[htb]\centering
  \includegraphics[scale=0.1]{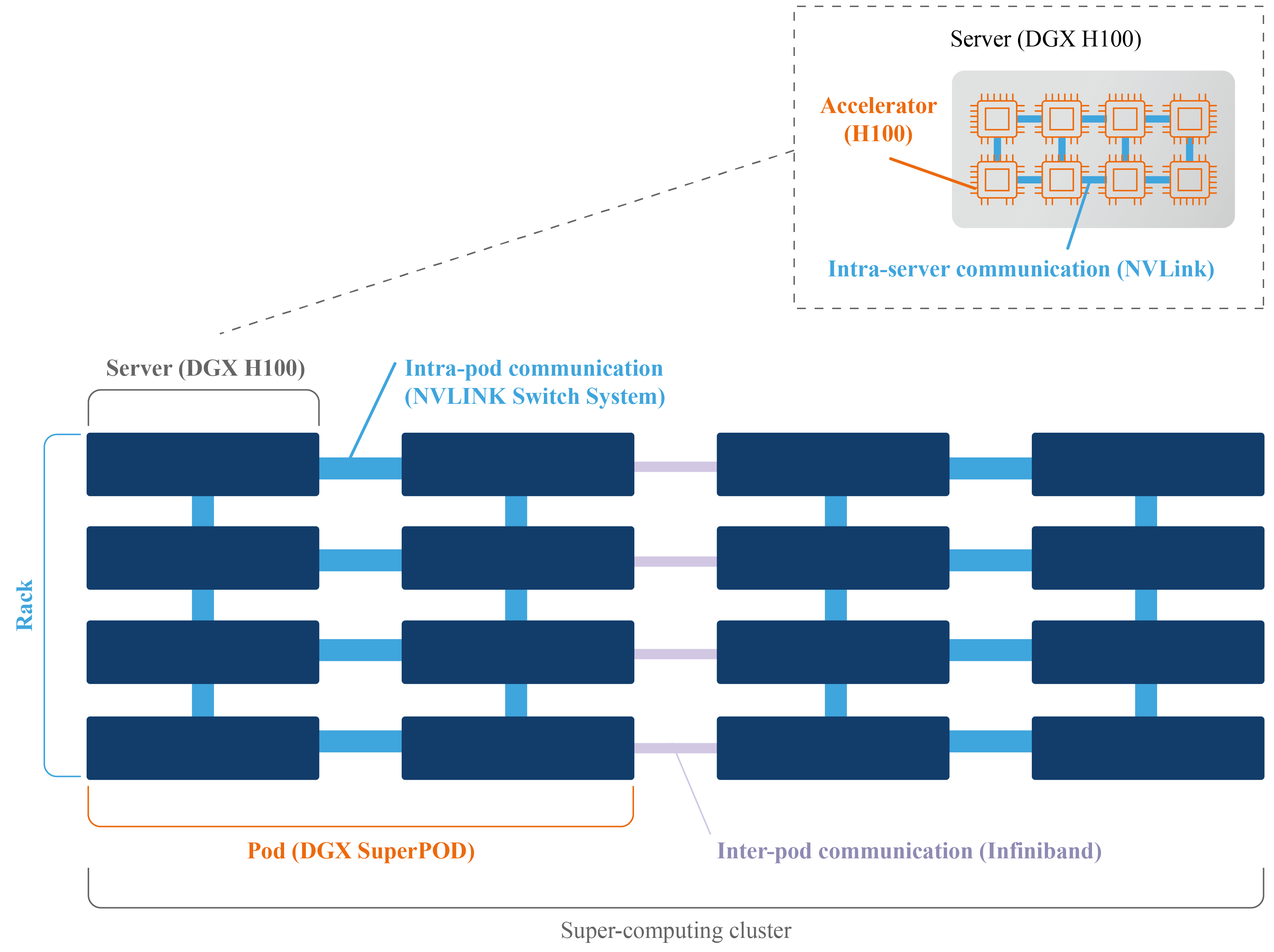}
\caption{AI Computing Cluster. Adapted from: \cite{Kulp2024}.}
\label{f6.png}
\end{figure}

\subsubsection{Related work}

\cite{Kulp2024} introduce the concept of a ``fixed set'' with a pre-configured set of AI chips and limited external communication bandwidth. \cite{petrie2024mechanisms}
discuss the potential role of encrypted interconnect in implementing controls on the number of interconnected GPUs.

\subsubsection{Technical details}

One approach to verify and enforce specific configurations of devices involved in an AI cluster is the \textbf{fixed set} approach. This involves assembling AI chips into an integrated pod, where each chip is pre-authorized to interconnect only with other chips within that pod. The integrity of the connections between the chips and resistance to tampering could be assured using a hardware Root of Trust. Chips should be able to perform tests to detect unauthorized changes and be disabled if relevant changes are detected, such as chip firmware modifications or connection of additional chips to a particular pod.

\begin{figure}[htb]\centering
  \includegraphics[scale=0.1]{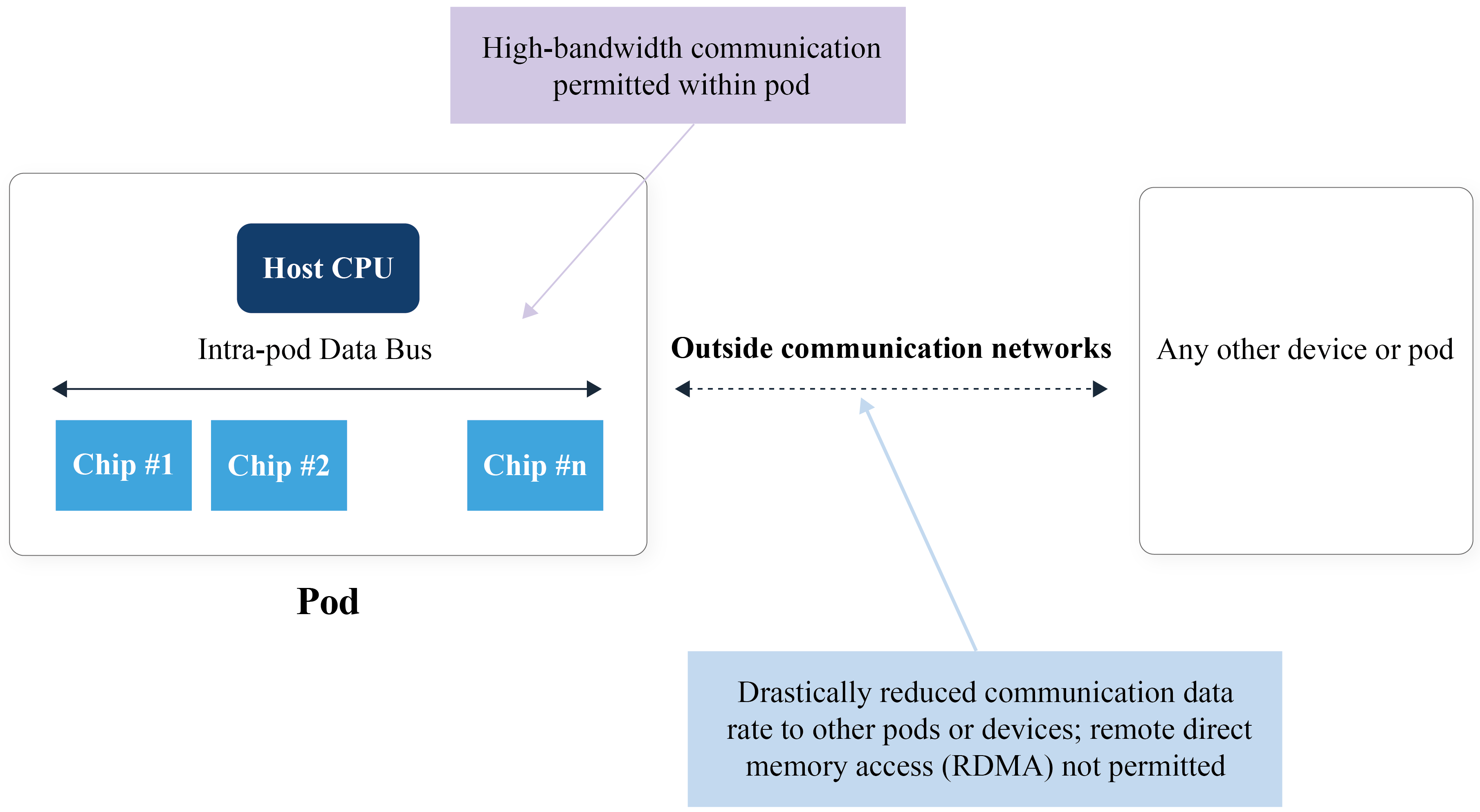}
\caption{Adapted from: \citep{Kulp2024}.}
\label{f7.png}
\end{figure}

A potential alternative is the \textbf{adjustable cap} approach, where each AI chip is equipped with a secure cryptographic ID, allowing it to authenticate the identity of other interconnected chips. Using public key cryptography or similar mechanisms, chips validate these IDs before exchanging data, ensuring that communication is limited to authorized chips. A cap on the number of chips that each AI chip can communicate with is enforced, which can be adjusted only with authorization from a trusted entity, such as a regulator. Chips periodically check this cap, and if the cap is lowered, they automatically limit or shut down communication with other chips. Similarly to the fixed set approach, to ensure the integrity of the mechanism it would be important that chips are able to detect unauthorized changes and that they no longer function if such changes are detected.

For both approaches, there is a question of what physical or software components within an AI cluster should be used to monitor the cluster configuration and enforce compliance. For a friendly actor, the easiest solution may involve management software, which can already expose rich information on a cluster's configuration to operators. Examples of such software for Nvidia products include NCCL \citep{NCCL2024NVIDIA}
at the framework level, UFM \citep{UFM2024NVIDIA}
at the networking level, or DCGM \citep{DCGM2024NVIDIA}
at the cluster level. However, it is unclear whether such software can be mandated and used to securely enforce restrictions without opening up low-cost opportunities for tampering or circumvention. In addition, data centers used for AI clusters employ different network topologies
\citep[e.g.][]{fattreeor2023From} and equipment from different suppliers which may have different associated management tools
\citep[e.g.][]{NvidiaBroadcom2024Frontier}. Given this heterogeneity, it is unclear how feasible it would be to adapt existing management systems to enforce restrictions across all possible combinations of hardware.

Alternatively, chip-to-chip interconnection could be limited at the level of individual equipment units and their firmware. This equipment could include either AI chips or networking equipment such as switches. This approach faces a different challenge: compute and networking components may face limitations on their ability to gather and interpret information about the cluster topology. For example, a host CPU in a server can see the local PCIe and network interfaces, but may lack information about the cluster's global topology. NICs and switches also only receive information about directly connected devices.

To address these issues, one potential solution could be encryption of all outbound interconnect traffic from high-performance AI chips. The provisioning of keys to decrypt this interconnect traffic could be used to limit the number of AI chips that can be interconnected. Encrypted interconnect could provide other benefits such as protecting weights from being exfiltrated by an attacker snooping on interconnect. It is also likely required for multi-GPU confidential computing, which can enhance the security of AI models against theft while they are in use on AI chips.

Nvidia's H100 chip already has dedicated AES-GCM hardware to encrypt PCIe traffic for single-GPU confidential computing. To enable multi-GPU confidential computing, AES-GCM hardware for the NVLink ports would likely also be required to ensure interconnect traffic is secure against interception. This suggests that future generations of AI chips may be equipped with the required encryption hardware to enable encrypted interconnect. However, if required hardware is not added, a potential alternative would be the use of occasional randomized authentication challenges, though this would likely provide lower levels of security.

More generally, a solution would need to be dynamic to account for swapping out units periodically, since the failure rate of AI chips (and networking equipment) in large clusters is non-negligible. This may be a particular challenge for the ``fixed set'' approach.

\subsubsection{Attacks}

Potential attacks on the integrity of these mechanisms could involve various strategies to bypass the restrictions on inter-chip communication. This could involve directly tampering with the mechanisms or circumventing restrictions to enable sufficient inter-chip communication for large-scale training.

In the \textbf{fixed pod} approach, adversaries might attempt to compromise the pod's Root of Trust or otherwise tamper with the system to connect additional unauthorized chips, allowing for larger-scale AI training beyond the preauthorized set. For the \textbf{adjustable cap} system, attackers could target the mechanism that enforces the cap on the number of interconnected chips, for example through physical tampering or attempting to broadcast instructions to the chip to raise the cap to levels not intended by authorized parties.

More generally, it may be possible to split the workload between supposedly separate clusters using data parallelization. If a small number of devices can be compromised, these could act as a bridge to smuggle in the averaged gradients between the separate clusters. If tampering to compromise devices is feasible but costly, this could significantly reduce the overall cost of circumventing restrictions compared to other mechanisms that would require every single device to be compromised.

To circumvent restrictions, data could be transferred indirectly via the PCIe connection to the CPU, using it as a bridge to other AI chips, instead of via direct chip-to-chip connections such as NVLink. This would likely result in a performance penalty relative to standard training approaches. To further deter this, it may be possible to increase the latency of PCIe connections.

\textbf{Decentralized training could present a larger challenge in future.} As decentralized AI training techniques improve \citep{peng2024decoupled, Patel2024Gigawatt, Prime2024INTELLECT}, attackers could exploit these advancements to use lower bandwidth between pods, making it feasible to coordinate large-scale AI training even when communication between pods is intentionally limited. Beyond the need to govern hardware at the level of a cluster, there may also be a need in future to govern systems of multiple clusters, which might even be situated in different countries. While the GPUs involved in training an AI model are usually located in a single data center or campus, there is strong interest from AI developers in performing training with a combination of several smaller clusters that are geographically separated.

This would increase the complexity and cost of implementing hardware-enabled governance mechanisms. For instance, if it is already a challenge to determine the boundaries of a single cluster, it may be yet more difficult to determine how many separate clusters are involved in a training run. Additionally, more distributed systems might vastly increase the number of entities that could contribute to training and whose hardware would be relevant to governance. Expanding the scope of monitoring and enforcement would exacerbate concerns about privacy and other impacts.

If decentralized training achieves sufficient performance to train AI systems with concerning capabilities, the greatest impact would likely be on mechanisms involved in workload attestation and cluster configuration limits, as it may be difficult to detect how large the effective cluster is and to establish broader, meaningful properties of its workload. Licensing and location attestation systems are less likely to be affected.

\subsubsection{Open research questions}
\begin{itemize}
\item What is the optimal mechanism of implementation for monitoring and enforcing chip interconnection limits? For example, how feasible is the use of cluster management software, network switches or individual AI chips for this purpose?

\item  How viable are secure, remote, post-manufacturing adjustments to previously specified limits on communication bandwidth? Can flexible caps or whitelists for interconnection of AI chips be implemented without creating security vulnerabilities that would make the system easy to bypass for moderately resourced actors?

\item  What are the appropriate technical parameters for pod size and external communication bandwidth limits, given current AI training needs and anticipated future developments in distributed AI training techniques?

\item  How can heterogeneous devices be identified in a secure way? How can individual device authentication or attestation mechanisms be integrated into a cohesive architecture that verifies and controls the number of interconnected AI chips?

\item  How can the integrity of fixed set pods be remotely attested, and what mechanisms could be developed to detect tampering with these configurations?

\item  If restrictions are enforced using networking equipment, how can this be done without undesired impacts on non-AI training activities that use this equipment?

\item  Is there a way to securely update which chips are permitted in the pod so that broken hardware can be replaced?
\end{itemize}

\subsection{Location verification}

AI chips could be designed to make it possible for an external regulatory authority to securely and reliably determine their location and take policy actions based on a chip’s location. Alternatively, the chip could determine its own location and respond directly, such as restricting operation when located in an area known to be subject to export controls. There are a number of technical approaches to location verification which can be used alone or in tandem to provide location estimates that are accurate and robust against adversarial tampering.

\subsubsection{Related work}

\cite{Brass2024Location}
provide an overview of the motivations and mechanisms for geolocation of AI chips for a general audience. There is an extensive research literature on the technical challenges of geolocation of devices in a non-AI context, which we discuss below.

\subsubsection{Technical details}

AI chips could participate in \textbf{challenge-response protocols} with landmark servers in known locations. Chips would establish an encrypted connection with a server over the internet, then receive and respond to a challenge sent by the server. The server would authenticate that the chip's response is signed by a private key unique to that chip, and would measure the time taken by the chip to respond to the challenge. This time delay could be converted into an estimate of the chip's distance from the server, and the chip's location could be triangulated using estimates of its distance from several landmark servers.

One technical issue with challenge-response protocols is estimating the chip’s distance from a landmark server based on the amount of time taken by the chip to respond to the server’s challenge. The relationship between response delay times over the Internet and distance can vary depending on how direct the routing path is, network congestion, or other factors. Some previous efforts have employed fixed ratios for converting delay times to distances \citep{maram2021goat}, while others have developed distance functions
\citep{laki2011spotter,padamanabban2001determining,sheng2024bft, kohls2022verloc,gueye2004constraint,wong2007octant}
to account for specific factors that drive variation in communication times.

Another technical challenge with challenge-response protocols is triangulating a chip's location based on estimates of distance to individual servers. A straightforward approach would be to trace circles around each landmark using the distance estimates, and intersect them to define a region in which the target is likely located
\citep{gueye2004constraint,wong2007octant}, but more complex approaches may improve accuracy. One alternative is to define location probability regions using historical data for each landmark, and return the location with the highest likelihood in the joint distribution of all landmarks \citep{laki2011spotter,padamanabban2001determining,arif2010geoweight}.
\cite{abdou2015cpv} select three landmarks defining the smallest possible triangle in which the target claims to be located and use trigonometric approaches to verify if the target is actually located inside the triangle. \cite{sheng2024bft} use trigonometry to estimate the maximal possible distance between the claimed location and actual location of the target in every possible direction for each landmark. They then combine these observations to trace a region of possible geolocation. Importantly, their protocol is robust to a percentage of compromised landmarks, providing Byzantine fault-tolerance.
\cite{kohls2022verloc} use gradient-descent based optimization techniques to find the most likely location based on a set of distance estimates from different landmarks.

\begin{figure}[htb]\centering
  \includegraphics[scale=0.1]{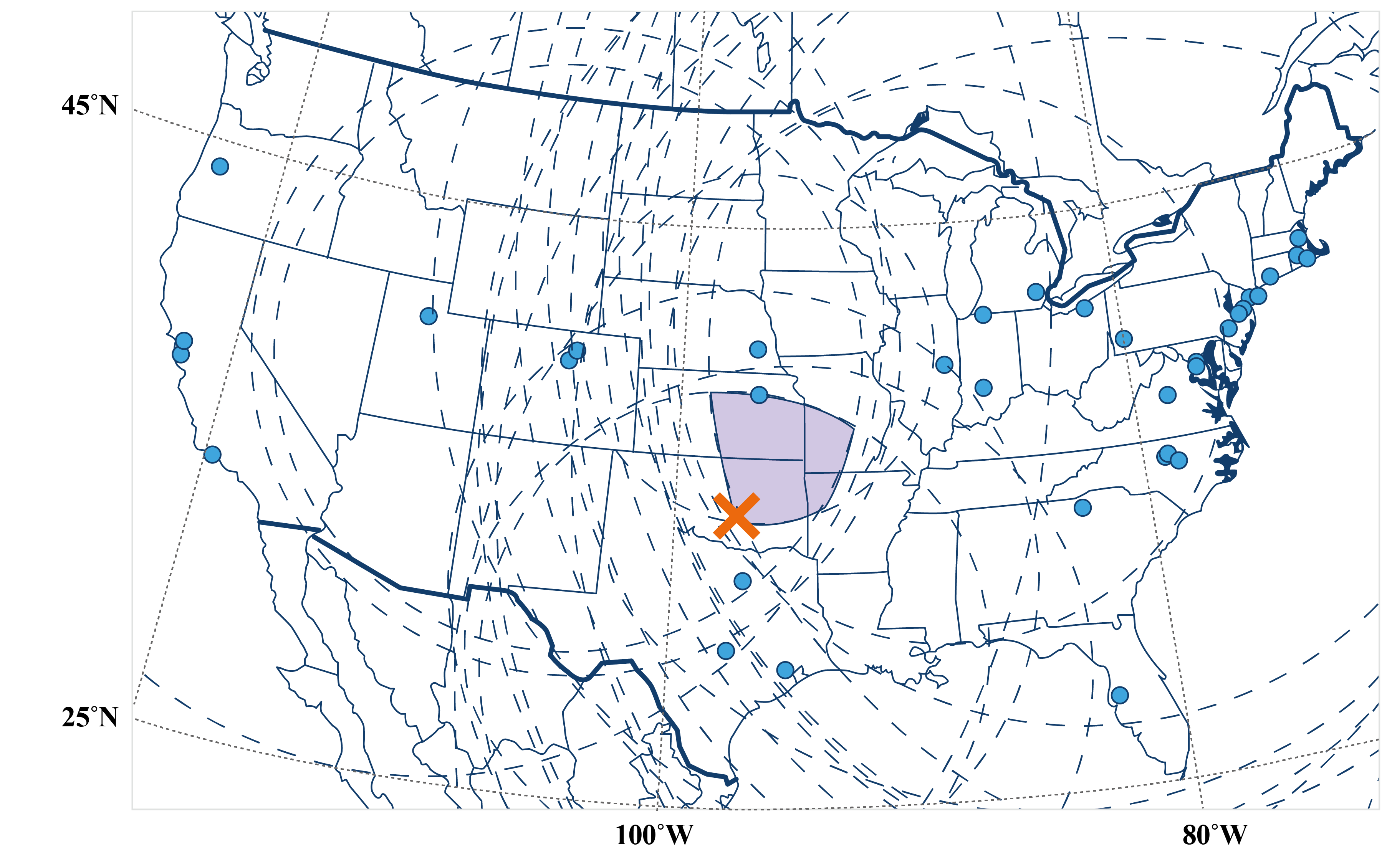}
\caption{By combining delay measurements from multiple landmarks, it is possible to estimate the location of an AI chip over the Internet or satellite networks. Adapted from: \cite{Gondree2013}.}
\label{f8.png}
\end{figure}

\textbf{Network infrastructure data} could provide additional validation of challenge-response geolocation results. Analysis of traceroute data and DNS records can help verify whether the network path taken by packets is consistent with measured distances. Further validation could come from consulting free or commercial Internet topology databases that contain IP address information. Internet Service Providers (ISPs) could also provide detailed network topology information, while chip users could report their ISP affiliations, offering additional cross-verification of the measured locations. These network-based validation methods, while not sufficient on their own, can serve as complementary verification mechanisms to strengthen the overall confidence in location estimates.

\textbf{Satellite systems} might be able to provide more secure and accurate geolocation by routing traffic from landmarks over satellite connections or even by locating landmarks on satellites. GPS systems are vulnerable to location spoofing attacks, but other services such as STL (Satellite Time and Location) and authenticated GNSS services like Galileo OS-NMA may provide greater protection against spoofing. This would likely require AI chips’ owners or users to set up antennas to connect to satellites, and the operators of the geolocation scheme to pay for access to relevant satellite services. In addition, to defend against certain types of attacks such as replay attacks, there may be a need for a secure clock near the chip to act as a reliable source of time measurements. Assuming that satellite connections can provide sufficient security and that the cost to use satellite infrastructure is manageable, these could serve as a complement or a substitute for internet-based challenge-response protocols. This could be particularly valuable in situations where unusually high levels of accuracy are required, such as for AI chip users that are located close to borders with countries subject to export controls.

\subsubsection{Attacks}

We highlight three main classes of attacks against challenge-response geolocation: manipulating response times, interfering with landmarks, and tampering with chips.

Response times could be manipulated to be slower or faster than normal. Slower responses would raise uncertainty about a chip’s location, but this might not be problematic, as regulators could simply identify this anomaly and assume the worst: if measurements cannot rule out that the chip is located in a restricted location, they could assume that it is. Faster response times would be more concerning, as this would make a chip appear closer to a landmark than it is in reality. These may be possible by using satellite connections or private terrestrial connections (for example by renting dark fiber) that can avoid potential delays due to circuitous routes or congestion affecting traffic sent over the Internet. To guard against this risk, many landmarks could take many measurements, making it difficult for an adversary to consistently spoof a single location. Alternatively, rather than manipulate the chip’s response times, the calibration of the landmarks could be manipulated to cause the regulator to consistently misidentify chip locations. Collusion between an adversary and an internet service provider could make these attacks easier.

Landmark servers could be manipulated in several ways. They could be subject to distributed denial of service (DDoS) attacks, rendering them unable to participate in geolocation protocols. Cyberattacks against a landmark's clock or timestamp system could cause it to misreport the amount of time taken for chips to respond. Attackers could attempt to locate landmark servers to inform their efforts to spoof a chip's location by speeding up and slowing down responses. Defenses against all of these attacks would involve having many landmark servers with strong cyber- and physical security that conduct repeated efforts at geolocation.

Owners of chips could tamper with their devices to undermine the cryptographic protocol. In this system, chips would sign their responses with a private key. If the chip’s owner can read that key and load the key onto another chip that can run the same protocol, then that chip could be placed in a different location and used to respond to challenges from landmark servers. This would effectively spoof the geolocation protocol. Success in challenge-response geolocation will require that chips contain private keys which cannot be read by chip owners.

\subsubsection{Open research questions}
\begin{itemize}
\item Given that the delay to distance relationship varies depending on local network conditions, what level of accuracy in location estimates is feasible in various regions of the world, particularly those in and around countries subject to export controls on AI chips?

\item  What improvements are possible on existing protocols for converting time delays to distances and distances from individual servers to absolute locations?

\item  How can location verification protocols be calibrated or complemented with other tools to minimize the rate of false positives, which could lead to unnecessary operational disruption for chip users?

\item  How scalable are the proposed location verification protocols, and what modifications might be required to enable location verification for millions of high-performance AI chips?

\item  Given their potential for greater accuracy, can satellite-based communications provide a sufficiently secure and economically feasible alternative to communication over the Internet for verifying locations?

\item  How can landmarks be secured against DDoS and other kinds of attacks?

\item  How can secure memory protect private keys from being read by chip owners?

\item  How can this protocol be made compatible with desires for strong cybersecurity at data centers that might involve airgapping?
\end{itemize}

\subsection{Offline Licensing}

In certain scenarios, AI developers or governments may wish to implement licensing regimes for AI accelerators. For example, when exporting AI chips to countries with a heightened risk of theft of AI chips or onward re-export towards export-controlled countries, it may be desirable to introduce licenses that limit the benefits of such activities by restricting the chip's functionality if it is stolen or diverted. More generally, this licensing mechanism would prevent the unlicensed use of AI chips, providing a flexible mechanism to monitor and control AI development and deployment in cases where the risks warrant such a scheme and where it is authorized by national regulation or corporate policies. Licenses could be implemented in the form of cryptographic keys that act as temporary passwords, unlocking a chip's capacity to perform a specified amount of computational work, such as a set number of operations or memory transfers. Once this computational work has been performed, the license would expire, and the chip would shut down or operate only at a reduced capacity. The chip operator would then need to acquire a new license from a license provider to resume full use of the chip. 

\subsubsection{Related work}

\cite{Kulp2024} proposed licensing as a mechanism for AI governance, and \cite{petrieII2024page19} detailed a specific implementation of a licensing mechanism. This licensing mechanism is closely related to Secure Boot \citep{SecureBoot2024}, a common security feature on personal computers that checks the cryptographic signature of a system's firmware to ensure authenticity before booting the system. One example of a licensing system that has already been deployed for commercial purposes is Intel On Demand \citep{IntelOnDemand2024}, which requires customers to purchase licenses to unlock certain chip functionality. There is plenty of existing work on secure boot in general, but research on hardware-based AI licensing mechanisms is currently nascent.

\subsubsection{Technical details}

Three core components of any licensing mechanism are licenses, meters, and throttling actions \citep{Kulp2024}.

\textit{Meters} would track the computational capabilities that are intended to be licensed and limited. Many different metrics could be metered, including ``wall-clock time, uptime, total energy intake (watt-hours), multiply-accumulate operations, amount of data transfer to memory (bytes) through the GPU's memory physical interface (PHY), or any other indicator of use that can serve as a workload distinguisher''
\citep{Kulp2024}. Many of these metrics are already tracked on modern GPUs. \cite{petrieII2024page19} argues that clock cycles are a good choice for a single metric to monitor, as they are relatively easy to track with lower risk of tampering and serve as an effective proxy for the number of operations used in training neural networks. Tracking multiple metrics might provide additional benefits, such as creating redundant measurements that reduce risks from tampering and providing flexibility to regulators in case the key inputs to AI development change over time. 

\begin{table}[htb]
\caption{Potential Targets of Metering. Adapted from: \cite{Kulp2024}.}
\label{t61}
\centering
\begin{tabular}{ll}\toprule
\textbf{Resource} &\textbf{Workload Target}\\\midrule
Floating-point arithmetic unit uses & High-accuracy matrix multiplication \\
Integer arithmetic unit uses & High-speed matrix multiplication \\
Memory transfer volume & Weights, activations, and dataset batches \\
Interconnect transfer volume (e.g., NVLink) & Distributed training, distributed inference \\
PCle transfer volume & Model checkpointing, distributed training, dataset batches \\
Joules (watts $\times$ time) & Hardware utilization \\
Clock cycles & Hardware utilization \\\bottomrule
\end{tabular}
\end{table}

\textit{Throttling actions} would limit or eliminate computational capabilities when a valid license is not present. These actions could throttle only one of the capabilities (e.g. memory transfers) while leaving others intact, or could limit all capabilities at once. Capabilities could be eliminated entirely or merely restricted to a reduced level. Secure Boot already implements an extreme form of throttling action, as the chip doesn’t boot if the firmware isn’t authentic. Further work is required to understand methods for circumventing throttling actions that are possible with current chips and for resisting these attacks.

\textit{Licenses} specify the amount of computational work that the owner of the license should be allowed to perform. This is a digital license which the chip itself reads and is unrelated to other forms of licensing such as export approval licenses for AI chips. A license provider could cryptographically sign their licenses with a private key and, if chips store the license provider's public key in unmodifiable memory, then chips would be able to distinguish authentic licenses from counterfeits. Licenses could also include other information, such as the ID of the license itself and an ID number for the device to which the license is intended to apply.

\subsubsection{Attacks}

We outline several categories of attacks on licenses and propose specific modifications to the licensing system design to thwart these attacks \citep{petrieII2024page19}:
\begin{itemize}
\item  First, to prevent chip operators from creating their own counterfeit licenses, the license provider could sign each license with their private key, and chips could be built to store the license provider's public key in secure, tamper-resistant memory, allowing the chip to verify that the license's signature is authentic.

\item  Second, to prevent the same license from being used multiple times on a single chip, the license could include a license ID that starts at zero for the first license issued to a particular chip operator and increments up with each new license. The chip could be designed to reject any license ID lower than the chip's most recently used license ID number, thus preventing old licenses from being reused.

\item  Third, to prevent the same license from being used for multiple chips, chips could be built with an unmodifiable Device ID, and licenses could specify the Device ID of the chip they're intended to license.
\end{itemize}

\begin{figure}[htb]\centering
  \includegraphics[scale=0.1]{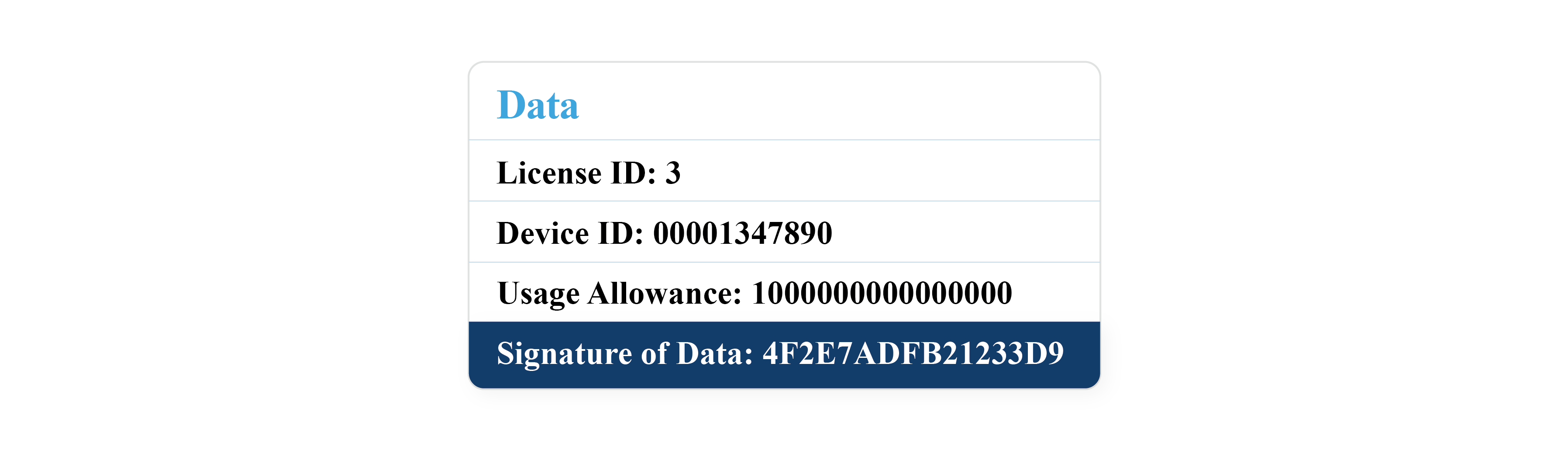}
\caption{Example License. Adapted from: \cite{petrieII2024page19}.}
\label{f5.png}
\end{figure}

The license itself is not the only component of the mechanism that could be attacked. Tampering with meters could cause inaccurate logging of chip usage, allowing the chip to be used beyond its licensed capacity. If the throttling mechanism can be circumvented, then the chip can similarly be used illicitly. Powering off a chip could enable mechanism tampering. More broadly, secure non-volatile memory is an essential component of the proposal in \cite{petrieII2024page19}, required to store the device ID and the license provider's public key.

\subsubsection{Open research questions}

Topics requiring further research related to this proposal include:
\begin{itemize}
\item  To what extent are existing secure boot technologies appropriate for protecting licensing (and other hardware-enabled governance mechanisms) against tampering? What new vulnerabilities could arise in this context and how can these be mitigated?

 \item How can the authenticity of licenses be verified in a scalable way across hundreds of thousands of chips? What other implementation challenges should be expected in deploying a licensing scheme at scale, and how can these be addressed?

 \item Which quantities should be metered, and how can this be done securely? What are the strengths and limitations of potential candidate metrics such as floating-point arithmetic unit uses?

\item  What technical and operational constraints should be considered when determining how much of each quantity should be allowed per license? Smaller limits would require chip owners to more frequently renew their licenses, which has costs and benefits from a governance perspective.

\item How should licenses be issued? This is primarily a policy question, not a technical question. However, technical researchers could enable more desirable policy choices, such as designing systems for multi-party provision of licenses that enable multilateral AI governance.\end{itemize}

\section{Challenges}
\subsection{Overview}

Developing and implementing the mechanisms discussed in this paper will present some important challenges, ranging from security measures to protect the integrity of their functionality to meeting operational constraints to ensure they can be deployed at scale. We will now outline some of the main obstacles that will need to be overcome to realize these mechanisms in practice.

\subsubsection{Physical tampering}

\textbf{Hardware-enabled mechanisms must be secure against potential adversaries.} Hardware-enabled mechanisms might provide the most value in adversarial environments where legal remedies or other existing approaches cannot be trusted. For example, if various national governments wish to make an agreement about frontier AI development, they will need trustworthy mechanisms for verifying compliance with the agreement that cannot be evaded by either side. Hardware-enabled mechanisms might play a special role here.

To characterize the different kinds of adversarial settings, each with different requirements for security against adversaries, we follow the taxonomy from \cite{aarne2024secure}:
\begin{enumerate}
\item  ``Minimally adversarial contexts, where attackers do not spend much on attacks, and are very averse to being discovered attempting to compromise mechanisms''

\item  ``Covertly adversarial contexts, where attackers are more willing to spend substantial resources to compromise mechanisms, but still want to avoid being caught doing so''

\item  ``Openly adversarial contexts, where attackers are willing to spend very significant resources to compromise mechanisms and are indifferent to this being discovered.''
\end{enumerate}

Physical security against tampering will need to be prioritized. Historically, one of the major motivations for work on chip security has been to preserve users’ privacy and confidentiality, by guarding against attacks that aim to infer secret information, hijack control flow, compromise system root-of-trust or steal intellectual property \citep{hu2020overview}. However, monitoring the activities of AI chips to ensure they do not violate regulations or agreements implies a different kind of threat model, in which the users themselves could be attackers. Under this new paradigm, anti-tampering measures will be essential to prevent people with physical access to GPUs from physically altering the mechanisms to circumvent restrictions.

\textbf{Anti-tampering measures may need an unusually high level of reliability in the context of hardware-enabled mechanisms.} Covertly evading compliance with the policies enforced by hardware-enabled mechanisms could potentially confer significant economic or military advantages. As a result, AI developers, governments or other actors may be strongly motivated to attempt to tamper with relevant AI hardware to disable or compromise these mechanisms. For regulators to trust the mechanisms, and for companies to adopt them, there will likely need to be strong evidence that they are robust and that it is sufficiently costly to circumvent them.

A defense in depth approach could improve stakeholders' confidence in the mechanisms' security; by layering multiple defense features, designers can reduce the likelihood of an adversary managing to bypass all of them and conduct a successful attack. These defenses could include destructive responses, such as hardware deactivation or data deletion.

System designs that reduce the attack surface could also be explored as a potential way of improving security and confidence. This might be done by running monitoring and enforcement operations on well-encrypted, trusted components like hardware security modules, for example, minimicing the number of components that need to be defended.

Formal verification of how software components function under different attack conditions could also provide strong evidence of their security. Although it would likely be unfeasible to obtain exhaustive proof for every single component under every relevant set of circumstances, it may nevertheless be valuable to pursue the most security-critical ones.

Finally, routine auditing for compliance with FIPS-inspired standards tailored to AI hardware could offer governments and operators additional reassurance of the mechanisms' security.

All these ideas require extensive research into the details of their implementation and the level of confidence they can provide, but they may be useful starting points for the development of the highly robust security needed. We will explore some of the technical requirements and concepts for possible anti-tamper measures in more detail later in this section.

\subsubsection{Privacy}

\textbf{Communication of data off-chip will need to be highly secure and limited in scope.} AI chips will need to share data beyond the chip for some of the mechanisms described above. Companies and chip users may be concerned that this could create new vulnerabilities, increasing the risk that their intellectual property or other sensitive data might be leaked or stolen. Measures will need to be taken to provide assurances to chip users regarding controls of which data can be shared beyond the chip, as well as to ensure strong information security practices on the part of those receiving any data.

If regulators have too much access to information about chip activities and too much power to restrict their usage, this centralization of control could bring its own risks \citep{sastry2024computing}. There may be concerns about misuse or overreach in surveillance and control, which could hinder legitimate commercial activities, restrict innovation or infringe on personal freedoms.

\subsubsection{Time, operational and cost constraints}

\textbf{Delay in the deployment of hardware-enabled governance mechanisms may significantly reduce their usefulness and impact. } Hardware-enabled mechanisms can help to address existing policy needs in areas such as verification of compliance with reporting requirements and export controls. They may play a critical role in enabling international agreements on responsible AI development. Further compounding the urgency of this issue is that chips without hardware-enabled mechanisms are currently being produced in increasingly large quantities, in response to rapidly growing demand \citep{TechInsights2024Nvidia}. The longer it takes to develop and implement, the larger the pool of chips without hardware-enabled mechanisms, which are more challenging to govern, will grow. Therefore, timeliness in the development of mechanisms that are ready for deployment at scale is of critical importance. In this context, approaches that require significant changes to the design of AI accelerators or other AI hardware appear less promising, as they could require several years to progress to production. Conversely, approaches that allow retrofitting to the existing stock of AI accelerators, or some subset of these, are especially valuable.

\textbf{Mechanisms should be implemented with minimal disruption to the operation of AI hardware. }Hardware-enabled mechanisms need to be usable in high performance AI settings. Hardware-enabled mechanisms should make allowance for the thermal properties of AI accelerators which result in stringent cooling requirements. This may present a particular challenge to the development of tamper-resistant enclosures. In addition, many frontier AI developers are seeking to improve their cybersecurity through measures such as airgapping \citep{nevo2024securing}, which hardware-enabled mechanisms should accommodate when possible. Sometimes, it will not be possible for hardware-enabled mechanism designers to accommodate all of the preferences of frontier AI developers. For example, anti-tampering solutions might include physical enclosures for chips that prevent developers from physically accessing chips to perform maintenance and address hardware failures. These costs may be worthwhile, but should be carefully considered and minimized.

False alerts from certain mechanisms or functionality, such as tamper-responsive enclosures, could interfere with training. Even if they are rare for any given chip, in a cluster of thousands of AI chips a small rate of false alerts could still present meaningful operational challenges. For chip designers and users to adopt the mechanisms, they will therefore need to be shown to be highly reliable and to have acceptable impact on operations.

\textbf{Hardware-enabled mechanisms must be economically viable to include.} Incorporating secure hardware-enabled mechanisms into chips will inevitably add costs, both through the additional research and testing that will be required and through the cost of any extra hardware or software components, if required. Leading-edge AI chips are already expensive, so it is plausible that these additional features would not lead to a large percentage increase in costs. Nevertheless, mechanisms will need to be economically viable to include over the long term for designers and users to adopt them.

\subsubsection{Adaptability to future technological changes}

\textbf{Mechanisms will need to be adaptable to a variety of designs and suppliers.} AI hardware designs are likely to continue to evolve over time. Novel designs that potentially diverge significantly from existing ones may be introduced, both from current market leaders and from other companies. While the market for AI accelerators is currently highly concentrated under Nvidia’s market leadership, the recent proliferation of chip design projects from large technology companies such as Amazon, Microsoft and Meta and AI developers such as OpenAI is part of a trend towards greater diversification and fragmentation
\citep{TheVerge2025Chip}. It is therefore important not to overfit any hardware-enabled mechanisms to the features of the most prevalent chips today and to maintain flexibility to ensure that they can be made compatible with future designs.

Techniques used for AI development will also likely continue to evolve over time, which could have important implications for AI governance
\citep{OpenAI2024OpenAI,gu2023mamba}. Changes in approaches to AI training and inference could result in shifts in how AI infrastructure is designed, which metrics are most relevant for regulators to be tracking, and in other areas \citep{heim2024governing}. This enhances the value of flexible mechanisms that can be adapted to a wide range of potential future AI chip designs or configurations of AI training devices.

\subsection{Physical Tampering}
\subsubsection{Motivation and definitions}

\textbf{For hardware-enabled mechanisms to be reliable, chips must be protected against successful tampering. }For the mechanisms discussed previously to achieve their aims, it is essential to guard against the possibility that adversaries could tamper with hardware, potentially altering or circumventing its functionality or copying a private key that they could use to report false information.

\textbf{Different measures have different levels of robustness to tampering. }While it may be difficult to obtain a perfect guarantee that hardware is impossible to tamper with, some measures represent stronger protections than others. For the purposes of this paper, we use the following definitions:
\begin{itemize}
\item  Anti-tamper: any measure that makes it more difficult to physically tamper with hardware.

\item  Tamper-proof: measures that make tampering impossible within the capabilities and budget of any relevant threat actor

\item  Tamper-resistant: measures that increase the cost of tampering beyond the budgets of some attackers but not all relevant ones

\item  Tamper-evident: measures that do not prevent tampering, but mean that an inspector can quickly and cheaply notice if an attack has happened.

\item  Tamper-respondent: measures taken in response to detected tampering, such as deleting keys if the enclosure is opened.
\end{itemize}

For some research domains, it will be useful if these terms are defined more rigorously. Future work should explore the range of tamper-related definitions more formally with an information processing perspective on confidentiality, integrity, availability, and the categories of goals of each involved actor.

\textbf{Some aspects of hardware attacks are becoming easier} \citep{hu2020overview}. As technology progresses, the technological means for attacks on hardware are becoming more sophisticated and more accessible. For example, scanning electron microscopes, which can be used to image a chip and extract information about its design and potential vulnerabilities, are getting cheaper. So too are focused ion beams, which are sometimes used in attempts to alter chips after they are already manufactured, by depositing or removing different materials. Side-channel techniques for obtaining information from hardware are also continually improving, with newer methods applying learning-based techniques to defeat established defenses
\citep{batina2019csi}. Additionally, the failure analysis methods required in semiconductor manufacturing provide a means of inspecting and modifying chip operation, but these same methods make it easier for attackers to do the same.

\textbf{Larger chips can make attacks harder to conduct, but also harder to detect.} As chip manufacturing evolves, individual features are getting smaller \citep{sun2020summarizing}. At the same time, the number of metal layers is increasing \citep{Srivatsa2020}. Other factors such as the adoption of advanced packaging add complexity to the physical construction of semiconductor devices \citep{Lau2022}. These developments can make it more difficult for attackers to find and target specific security features within chips. However, this same complexity can also make it harder for defenders to find Trojans hidden by attackers within chips \citep{bhunia2014hardware}.

\subsubsection{Research areas}

We will now outline some possible avenues for future research into protections against tampering.

\textbf{The security level of existing components needs to be established.} One important set of research questions relates to the security of existing components that might be used as part of the hardware-enabled mechanisms described in this report, including TEEs, TCMs, monotonic counters meters and non-volatile memory. Although these components are already in use, they are designed with different threat models than those assumed for our purposes. If they were incorporated into the previously discussed hardware-enabled mechanisms, they would become much more of a target for adversaries, and the potential harm of a successful attack on them could also be much higher. It is therefore essential to rigorously test these components, to properly understand their security properties and whether their affordances and features could be adapted successfully to the relevant threat models.

\textbf{Existing protections against tampering should be tested in the context of AI.} There are already many techniques to reduce the risk of hardware tampering, but they have not yet been extensively tested with the kinds of high-performance chips used in AI training. It would therefore be valuable to investigate how well and robustly these existing methods can be applied to the relevant feature size, power consumption, clock speeds, and packaging methods. However, this may be difficult to do without the cooperation of AI chip designers. In this case, FPGAs may be a useful alternative starting point for testing some relevant methods. The following are some examples of existing anti-tamper measures to investigate.

\textbf{Tamper-respondent memory.} One technique for guarding against hardware tampering is a memory component that can detect unauthorized attempts to access or alter it and quickly respond by wiping its content. Future research could explore how this tamper-proof memory concept might be incorporated into AI hardware.

\textbf{Sophisticated tamper-evident and anti-tamper packaging materials.} There may be tamper-evident methods of mounting components that make it obvious if they have been taken apart and put back together. Manufacturers could also explore embedding components in materials that are difficult to work with, which would make it more expensive for attackers to scale industrial processes to modify devices.

\textbf{Tamper-respondent enclosures.} Chips could be designed with multiple sensors to monitor electrical properties of their enclosures, such as capacitance and resistance \citep{Mosavirik2022}. Since tampering attempts could affect these properties, a chip could be designed to respond to anomalous readings by deactivating itself or wiping its memory.

\textbf{Stronger protections against tampering enable stricter threat models.} In addition to looking into how well existing techniques can be translated to AI chips, there is a research gap in promising concepts for stronger protections than those in use today. This is because much of the existing literature on hardware security focuses on different threat models and target hardware, such as low-power smart cards and microcontrollers \citep{gupta2021taxonomy,Rahimi2021}, whereas AI hardware draws hundreds of watts, operates at clock speeds of gigahertz, and employs advanced packaging methods like 2.5D integration of chiplets \citep{NvidiaH1002024}. Adapting secure enclosures and other anti-tamper defenses to this context may require substantial work.

\textbf{Enclosures and physical envelopes.} One line of research could be to adapt and improve existing anti-tamper hardware enclosures, whether at the level of individual devices, servers, or racks. Enhancements should aim both to increase enclosures' security to an appropriate level for AI hardware and to ensure they are compatible with AI chips. For example, since AI chips generate significantly more heat than traditional computational hardware, any enclosure for them must be able to dissipate heat without weakening the security boundary.

A potentially tamper-proof enclosure might involve a more sophisticated system that autonomously detects and reacts to any attempts to breach it. One potential detection design could be an active mesh that senses any chassis-intrusion attempts \citep{Immler2018BTREPIDBT}. Security standards similar to the Federal Information Processing Standards (FIPS) 140 Levels 3 and 4 could have extensions for side-channel protections and high-performance-specific requirements which AI hardware must comply with \citep{NIST2019}. Another possibility is volume integrity monitoring with an anti-tamper radio—an approach that involves broadcasting radio waves inside the enclosure and sensing how they reflect and propagate  \citep{staat2021anti}. A breach of the enclosure would likely change these properties, and thus be detected. These mechanisms serve as detection methods which could be tied to any response action, such as deleting sensitive information like model weights and cryptographic keys.

One option for a tamper response is zeroization, which could involve deleting the key required for the GPU to boot up. This is already employed in some commercially deployed hardware security modules
\citep{IBM2012}. It could also be possible to deactivate the chip, such as by blowing fuses or deleting critical calibration constants from physical input-output interfaces. Another path for deactivation is to physically damage the die, for example by fracturing the bulk silicon or thermally destroying the metal layers and active region. Thermal energy can be stored chemically in ``nanothermite'': a thin layer of nanometer-scale reactive particles applied to the surface of the die which can be electrically ignited with energy stored in an off-chip capacitor within the enclosure \citep{Sevely2022Developing}. Other options include use of high voltage which causes individual transistors to break \citep{Tada2021Design}, use of acid released onto silicon \citep{pandey2018self} or heat-expanding material that causes a chip to shatter \citep{pandey2022towards}.

\textbf{Offline power sources. }Many of the ideas outlined above rely on electronic mechanisms. This could be a weakness if safeguards were suspended whenever there was no power to the hardware, leaving it vulnerable to tampering. One way of addressing this might be to ensure a continuous power supply, including during shipment. There are precedents for the provision of special transportation conditions, such as constant refrigeration for goods that need to be kept at cold temperatures. Research could also explore batteries and capacitors as a backup source in the event of power failures while the chips are being deployed.

\textbf{Physical unclonable functions (PUFs).} Rather than depending on continuous active intrusion detection, the natural manufacturing variations between enclosures can form the basis of cryptographic key generation, such that any tampering would cause key derivation to fail. By using this key to encrypt information inside such a tamper-evident PUF enclosure, this information’s availability and confidentiality are tied to the physical integrity of its enclosure \citep{Immler2018BTREPIDBT}. During device operation, after the key is derived from the enclosure, the sensing circuitry can switch its function to active monitoring to delete the derived key and any decrypted data if tampering is detected. This style of tamper-evident PUF enclosure is distinct from SRAM-based PUFs which do not confer anti-tamper protections to the rest of the chip, nor do they enable battery-free protection \citep{ARM2019Anchoring}.

\textbf{Interlocking mechanisms.} To provide defense-in-depth, multiple anti-tamper and tamper-response methods can cross-check and cross-trigger each other. This makes offensive research more difficult since the system cannot be decomposed into independent subsystems to study separately.

\textbf{There needs to be system-level protection against tampering.} Many of the ideas discussed so far focus on safeguards for specific hardware components like sensors and memory. However, mechanisms aimed at supporting verifiable claims about chip usage will not just store data, but also communicate it between components within the chip, for example in making performance counters available to the trusted execution environment. It is therefore essential to ensure not only that components are secure in isolation, but also that data processing and communication are safeguarded.

We have so far mostly focused on tampering at the level of a chip itself. However, risks also exist at the connections between AI chips and other components within a larger system, and within the separable components of an AI accelerator (notably compute chiplets, memory chiplets, and the interposer which connects them). For example, even if an attacker does not breach or alter an AI chip, they may find weaknesses at its interfaces allowing disassembly and reassembly. For example, in November 2023, there were reports of assembly lines to reconfigure consumer graphics cards to be suitable for data center AI usage \citep{TechPowerUp2023Anchoring}. These sorts of misuses can be countered with anti-tamper defenses and component pairing (whereby components compare identifiers and refuse to work with recombinant configurations).

In general, holistic security involves considering different layers of abstraction independently and in aggregate. Some attacks, such as Spectre and Meltdown, come from leaky abstractions between layers
\citep{lipp2018meltdown,kocher2020spectre}. To avoid this category of error, the system must be considered across levels of abstraction and across combinations of subsystems. This sort of analysis can be handled through exhaustive testing or formal verification.

\textbf{Planned obsolescence and design diversity could reduce the risk of a serious attack.}
The longer that chips are functioning, the more time adversaries have to search for ways of bypassing their security features. Assuming that the current lifetime is around 3 years for AI accelerators or that upgrade cycles to a new generation of products happen every two years, there is an inbuilt planned obsolescence which can help with limiting the impact of attacks \citep{ostrouchov2020,META2024llama}. Assuming that vulnerabilities from previous generations of accelerators are patched in the next generation, ongoing efforts to breach mechanisms would be interrupted and attackers would have to start again.

Additionally, if AI chips have a wide variety of designs, for example by making key portions into reconfigurable logic fabric when performance and other constraints allow, then even if adversaries figured out how to successfully undermine one design, they would still only be able to attack a fraction of all chips. In these ways, planned obsolescence and design diversity can both reduce the likelihood that a successful attack will take place and limit the amount of harm that would result if it did.

\subsection{Privacy}
\subsubsection{Challenges}

\textbf{Hardware-enabled mechanisms can be designed with protections against government misuse.} A highly intrusive approach to oversight of AI development might involve government agencies having unrestricted access to AI models and information about what AI chips are being used for. This kind of approach would clearly offer a wide scope for abuse by monitors, and would not be appropriate. However, more narrowly targeted, carefully designed measures could strengthen transparency and confidence in responsible AI development practices, while addressing concerns about overreach.

\textbf{Reporting of AI activities could pose threats to legitimate privacy and IP interests.} A primary concern about any potential hardware-enabled mechanisms is privacy. Both individuals and companies may have concerns about the scope of observation of activities conducted with AI chips. Furthermore, the sharing of data with a third party such as a government agency might make it more vulnerable to being leaked \citep{heim2024governing}. Companies may also worry about third parties having access to information about an AI model itself, which might be critical to their intellectual property. The legitimate desire to protect trade secrets and prevent them from leaking might therefore make AI developers reluctant to accept this type of oversight.

\textbf{Privacy concerns are particularly pertinent to government and military activities. }Besides companies, governments and militaries are also developing AI tools. To reduce the risk of misuse, hardware-enabled mechanisms would presumably also need to be applied to AI chips being deployed in these contexts. However, information about government and military capabilities is likely to be particularly sensitive and crucial to keep confidential. Computing hardware is already a major target for espionage, and introducing such mechanisms to AI chips might create further vulnerabilities for the most highly confidential information to leak or be stolen by adversaries.

\textbf{Fears about leaking information could exacerbate pressures to race. }If hardware-enabled mechanisms leak information then this could accelerate AI development, because developers might learn about breakthroughs that their competitors have made and use these to enhance their own technology. Even just the fear of leaks could exacerbate racing dynamics; if AI developers feel less certain that their information is secure, then they might feel more pressure to accelerate their technology development in an attempt to keep ahead of potential leaks. Furthermore, any leaked information from competitors could also spur them to hasten development if they perceive that they are not ahead---or far enough ahead---of their rivals. The overall effect might be a more intensified AI race with reduced regard for safety measures, increasing the risk of dangerous AI models being released.

\textbf{Previous attempts to use hardware for surveillance have faced widespread criticism.}  In the 1990s, the US National Security Agency developed the so-called Clipper chip, which had a hardware backdoor built in to allow US government agencies to access voice messages if they established their authority to do so \citep{EPIC1993}. The government sought to introduce the chips in telecommunications systems, arguing that this would enable them to thwart terrorist activities. However, a combination of concerns, including individuals’ privacy and weaknesses in the technology that could lead to leaks, meant that the chips were never widely adopted.

This example illustrates the legitimate concerns of researchers and wider society about using hardware for surveillance. Since researchers and society at large are likely to be skeptical about proposals that sound similar, it is essential that any plans to monitor AI hardware should have robust protections built in against potential abuse by governments. Otherwise any proposed measures may also fail to be implemented.

\textbf{Centralizing power could increase risks of government misuse.} While monitoring AI chips seeks to reduce the risk of them being used for harmful purposes, this approach may also grant too high a level of control to governments and government bodies. This centralization introduces different but equally serious risks.

If a government body is monitoring AI chips, it could have extensive access to information about companies and the population at large, and potentially also control over which AI activities are carried out. This could create the potential for a government to increase its surveillance of and control over the public, beyond what is reasonable to prevent harmful activities by individual or corporate actors. It is worth noting, however, that there are already many other tools for digital surveillance used by authoritarian governments today, such as facial recognition and internet censorship. While hardware-enabled mechanisms might be misused to support this kind of control, the marginal impact relative to existing digital surveillance activities is unclear.

\textbf{Centralizing power could increase the potential harms of malicious attacks.} Even if the monitoring agency itself did not seek to misuse its access to sensitive information or its control over activities, simply having the power to do so would likely make it a target for malicious actors. Such actors might be able to leverage reporting mechanisms to access sensitive information or misuse enforcement mechanisms to disrupt legitimate AI activities.

We will now explore some of the potential ways that monitoring mechanisms could be designed to mitigate privacy and centralization concerns, while still reducing the risk of harm by monitoring the use of AI chips.

\subsubsection{Mitigations}

\textbf{Only the highest-risk hardware should be monitored. }A first step towards protecting privacy could be to ensure that any hardware-enabled mechanisms are applied only to the most relevant hardware. Since some types of hardware have less potential to be used for harm, even in the hands of a malicious user, they could be excluded from the measures. For example, non-AI chips and some AI chips like gaming GPUs, which are not designed to be used in data centers or for training large-scale AI models, could be exempted.

Large-scale systems of high-performance AI chips would pose the greatest risk if they fell into the wrong hands, but they currently make up only a tiny fraction of all chips produced, and are prohibitively expensive to individuals and even most companies. Restricting any monitoring measures to these chips would leave the vast majority of computational resources unmonitored, reducing the concern of widespread public surveillance.

It is worth noting, however, that as technology progresses, it may become possible to train dangerous AI models on cheaper, more accessible hardware, expanding the number of chips that might be considered to require monitoring. Any regulations would therefore need to be reviewed and adapted, with appropriate consideration of the changing landscape.

\textbf{Monitoring might only be appropriate in the case of catastrophic risks.} Expanding on the concept of reserving monitoring for only the highest-risk hardware, it could be argued that it would only make sense to apply it where there is a risk of a catastrophic event or a risk to national security. The rationale for this argument is that monitoring is a form of preventative measure, and could therefore be limited to hardware with the potential to be used for harms so great that no subsequent penalty against the developer would suffice. For hardware with the potential to be used for lesser harms, regulations could rely instead on imposing penalties against developers if an adverse event were to actually happen, rather than monitoring to pre-empt such an event. Following this principle could help to narrowly target any monitoring activities, for example by focusing on models that could be used to develop novel weapons or deadly pathogens, helping to address concerns about broad surveillance.

\textbf{Privacy-protecting principles should be incorporated into hardware-enabled mechanisms.} To protect individuals' privacy, any mechanisms should be designed to disclose only highly specified data that can be used to assess a model's capabilities, without sharing any information beyond that. Particular attention should be paid to ensuring that sensitive information is not shared. Since these mechanisms unavoidably introduce additional communications channels, care should be taken to ensure these channels are highly secure, to protect against leaks. Any information should be shared with the minimum number of people necessary.

\textbf{There are promising technical approaches to balancing monitoring and privacy.} For example, trusted execution environments could facilitate privacy-preserving evaluation and auditing. Chips with in-built trusted execution environments (TEEs) might allow an auditor to test whether an AI model complies with relevant standards, without actually being able to extract any critical information about the model, such as its weights. TEEs may also be used to facilitate privacy-preserving attestation of workload properties, as discussed in detail in the ``Workload Attestation'' section.

\textbf{Hardware-enabled mechanisms and related policies should be frequently reviewed.} As technology advances and access to it expands, regulations may become inadequate or inappropriate over time, with respect to both effectiveness at preventing harm and threats posed to privacy. For example, if it becomes possible and practical to train AI models with dangerous capabilities on consumer devices, many of the approaches described in this report would no longer be sufficient. Since the AI landscape is rapidly evolving, regulations should be frequently reviewed, at least annually, to keep up with these dynamics. Such reviews would also offer an opportunity not only to check whether the regulations had become outdated, but also more generally whether they had been effective up to that point and whether they could be adapted to be more so.

\textbf{All measures should be well defined with rigorous safeguards to prevent overreach.} To balance the need for oversight with protections for civil liberties, policymakers should incorporate robust safeguards against government overreach. There should be clear definitions of the scope of any measures, as well as who is allowed to access what information and under what circumstances. There should also be clear protocols surrounding the actions that officials may take to access information, and what actions they can take in response to suspected breaches of regulations. These boundaries could be defined by governments as legal limits, or even as limits to their powers set at the constitutional level.

Procedural safeguards can also play an important role in designing appropriate protections \citep{sastry2024computing}. These include measures like rulemaking processes that involve public input, protections for whistleblowers, roles for internal oversight and consumer advocacy within regulatory bodies, avenues for judicial review, advisory boards, and transparency through public reporting on activities.

\textbf{International governance could prevent misuse of powers.} To ensure that governments abide by the limits they set on regulatory powers, international processes could be developed as a form of mutual oversight. International agreements could state that controls should only be implemented where genuinely required for national security. As an ongoing measure, any update to hardware-enabled mechanisms could require agreement from multiple international parties to be implemented, with chips otherwise rejecting it. This might reduce the capacity for a single government to misuse available tools for expanded surveillance.

\textbf{Mechanisms should focus on permission-based sharing of data.} To increase trust, any disclosure of information should ideally only happen with the knowledge and consent of the hardware owner or operator. If AI developers are always aware of which information is being shared, so that nothing can be accessed without their knowledge, this may increase their willingness to cooperate while deterring regulators from overreaching.

\textbf{Rules could be enforced locally on-chip, reducing the need for data sharing.} As an alternative to chips sharing information with regulators, chips could instead be configured in such a way that they cannot be used for any activities that do not comply with rules that have been agreed upon by multiple governing parties. This could reduce the need for monitoring and reduce the concerns around data sharing and potential leaks. Another advantage is that it would serve as a preventative measure, aiming to avoid misuse occurring, rather than regulating hardware by detecting instances of rules being broken and punishing non-compliers afterwards.

\textbf{Open-source mechanisms could increase transparency and trust.} The design of hardware-enabled mechanisms could be made fully open source, so that AI developers can understand exactly what they do and how they work. In this way, developers could become more confident that the mechanisms were not going to allow for data to be shared without their knowledge. This kind of transparency should improve trust in the system and willingness to cooperate.

However, AI chip designers may be reasonably concerned that making mechanisms open source could mean sharing information about the hardware too widely, potentially making it easier for people to steal chip designers' intellectual property or circumvent chips' security features. For this reason, any open-sourcing would need to be done carefully, only including information about how the mechanism works, and nothing about the chip's design beyond that.

\section{Components}

The mechanisms proposed for governing AI hardware will only work if the chip-level components involved are functional and secure against attacks. Many types of components that could play a role in regulating AI are already used in traditional computing. However, more research is required to understand how they can be adapted for the specific use case of AI governance, as well as how they can be protected against tampering and common attacks to circumvent restrictions. Alongside security considerations, researchers must also be careful to ensure that designs for these components will not cause unintended consequences or interfere with how well they operate in AI training. We will now look at several components of interest one by one, detailing how they might be useful, the technical requirements for their use in AI governance, and some possible design options for meeting those needs. We will highlight some of the main open research questions surrounding the practical implementation of AI hardware governance.

\subsection{TCMs}

Trusted compute/cryptography modules (TCMs), also known as secure cryptoprocessors, are discrete components that are used to perform cryptographic operations. They enable the secure storage of a cryptographic key in one specific location on a chip, and, as discrete components, are less integrally connected with the rest of the chip. This separation makes it more difficult to steal the key through vulnerabilities in the main application processor such as microarchitectural side-channels. Additionally, since TCMs are relatively small and simple, they tend to be more secure and less likely to have hidden vulnerabilities.

TCMs are similar to the next component described below, the trusted execution environment, except that they are discrete (even when sharing the same die) and have a smaller set of functions like encryption, decryption, signing, and signature checks. A trusted execution environment, on the other hand, could include a TCM in its implementation to support the secure execution of arbitrary programs.

It is important to note that, although incorporating TCMs can increase security, they are not a complete solution by themselves. The operations of the TCM itself may be highly trustworthy, but it is still important to ensure that the information going into it can also be trusted; if the input has been corrupted in some way, then the output cannot be trusted even if it has been processed correctly.

If TCMs are to be used to facilitate hardware-enabled mechanisms, it is essential that they are secured against potential attempts to extract information from them. In particular, it is important to protect them against physical tampering and side-channel analysis, to reduce the likelihood of someone extracting the key.

Research analyzing potential attack methods and developing measures to reduce the risk of successful attacks is therefore needed. Formal verification of TCMs would ensure that some known classes of attacks are impossible. This level of certainty may be difficult to achieve, but one of the prime advantages of using a simpler component is that it is feasible to achieve stronger confidence in its security.

\subsection{TEEs}

Trusted execution environments (TEEs) are hardware features that guarantee a level of confidentiality and integrity to certain processes running in them. In their current form, TEEs allow cloud providers to offer confidential computing, whereby they can run software for customers without being able to access those users' data. While other components can store or transmit data securely, TEEs secure data while it is being processed.

These components could be adapted for multiple aspects of AI hardware governance, including running confidential tests to evaluate the properties of a model, and verifying certain quantities associated with the operation of the chip itself, such as the number of arithmetic operations conducted over a period of time \citep{aarne2024secure, reuel2024open}. In both cases, using TEEs should give regulators confidence that the measurements obtained are authentic, without assuming that hardware operators will voluntarily provide truthful reports.

In the context of model evaluations, this feature preserves confidentiality by ensuring that the regulator cannot directly access information about the model, and the hardware operator cannot directly access information about the tests being run. This is valuable to protect commercial secrets of model developers, and to prevent model developers from gaming the evaluations.

While this is an interesting research avenue, it could be that simpler mechanisms would be sufficient, such as using TCMs for cryptography and using secure meters to guarantee authentic counts of arithmetic operations. Additionally, since TEEs may require more encryption steps, they may introduce unacceptable latency to AI workloads in some cases \citep{zhu2024confidential}. As an alternative to using a TEE incorporated into the AI chip, it might be possible to instead use a separate component external to the AI chip, perhaps similar to the model used in the Intel Management Engine \citep{intel2022}. This approach might have the added benefit of leaving the chip’s performance unhindered; integrated TEEs require encryption as well as communication with other parts of the chip, imposing additional operations on the execution flow that could cause the chip to run more slowly; rather, keeping trusted execution resources separate from normal execution resources would allow a consistent interface for application code, provide a clear trust boundary, and not impact performance. It is unclear how much access the secure processor needs to the rest of the chip's state to have sufficient confidence in tasks like workload classification. For example, it might be sufficient to provide access to performance counters, or it might instead require more holistic access, such as the ability to introspect application VRAM. The more resources must be accessible to monitoring, the higher the risk of unintended loss of confidentiality and the larger the attack surface area.

In addition to model evaluations and compute accounting, TEEs could be used for location verification by signing and returning unique ping messages, allowing a landmark server to measure the round-trip time between sending the ping and receiving the response \citep{Brass2024Location}. The TEE provides for verifying authenticity of ping responses, such that spoofing the location to be farther than measured would require extracting the TEE's private key.

For these mechanisms to be effective, it is essential that the private key is stored only in the TEE, so that results cannot be forged from the copy of the key. The public portion of the key pair should be stored in a registry, so that it can be used to interact with the chip and verify its attestations against an ownership record. For the strongest security, the key pair should be generated on-device as opposed to a less secure approach where it is generated externally, copied in, and then deleted from the external source.

Research into tamper-proofing is essential to develop TEEs that are appropriately secure for the threat models applicable to hardware-enabled mechanisms, and are not vulnerable to attacks by actors with unsupervised physical access to them. The level of  physical security that such devices currently have is not designed to address threats from untrusted hardware operators or inspection-free operation.

Existing TEEs have security vulnerabilities that might lead to information leaking \citep{Munoz2023}. While the primary requirement for hardware-enabled mechanisms is to guarantee that results are authentic, information confidentiality is also a concern. Research into improving logical security could investigate designs that ensure TEEs do not share resources like cache with the rest of the chip, since these are a common attack vector through which adversaries try to obtain confidential information. More ambitiously, pursuing formal verification of the TEE could provide stronger security guarantees. Finally, chips which only allow execution in the TEE (without also supporting simultaneous execution in the untrusted context) would remove the need for separation within the chip between trusted and untrusted execution contexts.

Since existing TEEs are not designed specifically for monitoring purposes, research into the exact roles they could play within this sphere would be valuable. For example, if secure meters are used to ensure that measurements are not interfered with, making those metrics available to programs running in TEEs could offer a guarantee that those metrics are coming directly from the chips in question. Additionally, since training and deploying state of the art AI models involve large numbers of GPUs, trusted outputs from a TEE on just one chip might not give auditors enough useful information about whether aggregate activities breach regulations or not. It would therefore be useful for researchers to explore how networks of TEEs across many GPUs could report authenticated information about the model more broadly, rather than only at the level of a single chip.

\subsection{Monotonic counters}

Monotonic counters are components that can increase but not decrease in value when measuring activities being carried out on a chip. Crucially, they store the same state when the chip is switched off, so that this cannot be used to disrupt measurements.

In the context of hardware-enabled mechanisms, monotonic counters could keep track of various relevant properties, such as how many FLOPs have run on a chip. They could be used within mechanisms that deactivate the chip if it surpasses a threshold of allowed activities, or require it to request a new license when it reaches an upper limit. Monotonic counters could also facilitate licensing regimes by tracking how many licenses have been activated; for licenses which include a count of how many came before, this sequence number can be compared to a monotonic counter to notice re-activations and out-of-order activations of licenses. In general, any relevant increasing properties that must be recorded across reboots, such as the total number of arithmetic operations ever performed by the chip, should be stored in monotonic counters.

For monotonic counters to be useful, they must be guaranteed to only increase, and they must be made tamper-resistant so that they cannot be artificially lowered. Nonvolatile flash memory is a likely candidate to preserve the counter’s value when power is lost. Designs should consider adversarial timing of power loss which could result in improperly updated persistent storage. One setup could involve storing the value in SRAM, but including a capacitor that recognizes when the chip is about to switch off and reacts by immediately copying the value from SRAM to flash. Alternatively, in a setup without a capacitor, the counter value could be updated in flash memory every time it is incremented, or rounded up by a certain amount each time the chip boots up.

Although the latter approach would not need a capacitor, it would not be useful for keeping track of quantities that increase very quickly, such as an exact count of the total arithmetic operations performed. This is because it takes a relatively long time to save a value to flash memory and updating the flash too frequently could also wear it out. As a compromise, the chip could be configured to back up the value to flash at regular intervals, such as once per hour, with a mechanism involving a capacitor to save the current value to flash in the event of power loss.

To guarantee that a counter's value cannot be artificially reduced, it may be worth exploring unary counting, in which fuses are burned sequentially to represent each increment of the tracked value. This technique is likely only practical for counting properties that are incremented infrequently, and which are unlikely to reach a very high value, such as license activations.

\subsection{Secure meters}

Performance counters already exist as a standard feature in AI chips used to count operations including memory accesses, floating point operations, and interconnect data movement and for the purposes of debugging and profiling performance. Like performance counters, secure meters could be used to monitor and report AI chip properties. However, performance counters can be directly controlled and reset by the user, which should not be allowed for secure meters involved in hardware-enabled mechanisms.

Secure meters could be used for licensing and workload attestation by offering a way of collecting trustworthy information about activities conducted on a device. They could also be a practical way of implementing AI cluster configuration limits. For instance, in a scenario where certain bandwidth limits have been agreed, regulators could use secure meters to monitor the bandwidth of data being sent between chips to verify that the cluster does not exceed the bandwidth allowed for a particular user. Additionally, this information could be used to assess whether a user is complying with regulations, and adjust their connectivity cap accordingly.

In terms of technical requirements, secure meters should be able to count properties that are useful for AI monitoring, including memory transfer volume, off-chip transfer volume, arithmetic operation count, total instruction count, and tensor core usage. The secure meters' values should be outside users' control and impossible to reset except under allowed circumstances, such as upon license renewal. They should also be able to work in conjunction with other parts of the chip as necessary, such as TEEs, which might need to securely access their values to provide guarantees that the measurements have come from the correct chip and have not been tampered with.

Useful research avenues could explore how to adapt performance counters to monitor properties of interest, how to robustly prevent resetting or wiping of the counter under most circumstances, and how to integrate them within chips to ensure secure transmission of counters' outputs and protect against subsequent manipulation or corruption of measurements.

\subsection{Secure memory}

To support hardware-enabled mechanisms, AI chips need to be able to store persistent values that do not disappear when the chip loses power and are robustly protected against attacks aiming to alter them. Ideally, chips should be designed so that only a specific intended component of the chip, such as a TEE, can read and write values stored in relevant memory components. This level of separation from other parts of the chip might be achieved through encryption or through structural features that mean this secure memory is not connected or accessible except to one component. These requirements are nearly met by standard nonvolatile configuration storage in a system on chip (SoC) design, though with differences including more frequent rewriting and stricter security.

Many hardware-enabled mechanisms rely on the secure and persistent storage of a range of different types of information, so secure memory will be an essential enabling technology for AI governance. For example, it will be needed to store critical data including cryptographic keys and licenses, so that they are protected from manipulation and can thus be certified as valid. Secure memory could also be used to track dynamic quantities, such as compute graphs to build a record of the kind of computation being performed, and provide trust that it has not been forged or tampered with.

Research efforts should explore how to make secure memory reliably persistent, and how best to defend its integrity. As with all other components, it should also be made resistant to side-channel attacks and tampering. One potential route to achieving these properties might be to put standard flash memory inside a secure enclosure and filter its power supply, to guard against fault injection. Additionally, on-chip anti-tamper interventions could include detecting and responding to fault injection.

\subsection{RTCs}

Real-time clocks (RTCs) that are calibrated to the actual time, rather than the amount of time since the chip was last booted or its total number of clock cycles, could be useful components in hardware-enabled mechanisms. For example, they could ensure license expiration after a specific period of time, to prevent people from stockpiling licenses and then using them all at once, which might be a way of circumventing regulations. RTCs are a form of monotonic counter, and could similarly be used to ensure licenses are activated in the intended sequence; they have the additional advantage of enabling licenses to specify a point in time after which they can no longer be activated.

The primary technical requirements for RTCs are that they should keep time accurately and be resistant to attempts to change their value or functioning. A possible option for an RTC design is a secure monotonic counter that increments at a rate that is continuously actively calibrated for changes in temperature and pressure. Research should explore how such a component could be secured against attempts to slow down or speed up the rate at which it is incremented, for example by changing environmental conditions or interfering with the power supply. Other considerations could include tamper-response features that trigger certain protective actions if the sensors involved in calibration report an anomaly. These actions might include automatically deleting private keys used for licensing or chip identification, for example.

\subsection{Tamper proofing}

As mentioned throughout this section, tamper proofing will be essential for every component involved in hardware-enabled mechanisms, to guard against attacks leveraging physical access to the chips. Tamper proofing could take many forms, including designing existing components to add anti-tamper features or using additional protective components, such as secure enclosures. While physical, on-site inspections (potentially supplemented by video surveillance, which is already standard in many data centers, or tamper-evident seals) could go some way toward detecting tampering attempts and rectifying them, there is clearly a limit to how thorough such inspections could be. Ideally, preventative features should stop any tampering attacks from happening in the first place, and reduce the need to ensure that everyone with physical access to the hardware is trustworthy.

Anti-tamper features range from measures that make it more difficult to physically tamper with chips to mechanisms that detect such attempts to systems that respond to tampering detection, or a combination of these functions. Separately, another design objective is to make the enclosure tamper-evident: making it difficult to tamper with the device without leaving evidence which an inspector could easily notice, but the attacker cannot easily hide.

Approaches to automatically detecting tampering are discussed in the Physical Tampering section above. Additionally, countermeasures against side-channel attacks should be explored to make it sufficiently difficult to steal information from chips by inferring it from observed patterns in power consumption, electromagnetic fields, acoustics, etc. Along with these measures to prevent and detect tampering, we also need ways to respond to detected attempts. Such a response would serve to protect information and functionality from improper use. Protecting information is generally easier since it can be deleted, whereas destroying functionality is more difficult. One method is to physically remove it, such as by blowing fuses which encode critical information like calibration constants; another is to trigger a separate thermal destruction, such as by igniting a nanothermite layer; one more approach is to move functionality into the information domain by using field-programmable gate array fabric for sensitive intellectual property and deleting its configuration as a response.

Finally, physical unclonable functions (PUFs) should be explored as a possible means of removing the complication of battery-backed tamper-response systems. In this approach, critical information is encrypted with a key which is generated from measurements of the enclosure itself---if the enclosure is tampered with, these measurements will have a different outcome, thus tying the availability of the key to the physical integrity of the enclosure. Such an enclosure provides protection while powered off and could also be actively probed when turned on after key derivation to notice tamper attempts.

\subsection{Offline power source}

For active monitoring approaches, continuous function requires continuous power. To match current logistics methods, a reasonable solution is to include a battery within the enclosure to maintain operation even during reboots, shipping, and shelf time.

Research into offline power sources should investigate possible systems for reliably supplying power during shipment, similar to requirements for refrigerated shipping for frozen goods. Other possibilities to be explored include ways of protecting batteries from being removed, and fail-safes for complete power loss that trigger responses similar to those when tampering is detected.

Components that might be useful within these systems and mechanisms include various kinds of batteries, rechargeable or otherwise, and capacitors. Research into the most secure configurations could look at deploying multiple layers of batteries, wherein the removal of larger batteries outside the security perimeter could trigger a response by smaller batteries inside it after a delay suitable for maintenance of the external systems. Capacitors could be incorporated into mechanisms that notice when batteries are close to running out of energy to trigger a response even if the battery loses charge faster than expected.

\clearpage

\section{Conclusion}

Hardware-enabled mechanisms offer significant potential for advancing the governance of AI systems by enhancing regulatory visibility and improving compliance with safety standards while safeguarding intellectual property. This paper has outlined several mechanisms that appear especially promising in this regard. Verifiable training and inference mechanisms could allow external auditors or other third parties to confirm key properties of AI workloads — such as usage of computation — without requiring access to the model itself, preserving confidentiality while enabling accountability. Verifiable cluster configuration mechanisms could support claims made about which devices were involved in training or deploying an AI system and could be used to enforce future agreements or regulations on scaling of compute clusters. Location verification mechanisms could help enforce export controls with lower compliance costs and enable hardware to be used only in trusted jurisdictions. Lastly, offline licensing could allow hardware functionality to be throttled under conditions pre-specified by regulators or AI developers, supporting more granular oversight and control.

These tools could play a pivotal role in ensuring responsible AI development and fostering international cooperation, particularly as the capabilities and risks of advanced AI systems continue to grow. However, realizing this potential requires addressing several critical open questions. These include assessing the resilience of potential implementations against adversarial attacks and navigating potential trade-offs between security, privacy, and operational scalability. Implementation will depend on fostering collaboration between researchers, policy-makers and industry stakeholders to ensure mechanisms are effective and can be deployed at scale. By prioritizing these areas for further investigation, researchers can unlock the full promise of hardware-enabled mechanisms as tools for ensuring secure and responsible AI development.

\bibliographystyle{apa-good}
\bibliography{references}

\section*{Authors}

\textbf{Aidan O'Gara} is a PhD student in AI Governance at Oxford University and AI Program Officer at Longview Philanthropy.

\textbf{Gabriel Kulp} is a Technology and Security Policy Fellow at RAND and a PhD student at Oregon State University.

\textbf{Will Hodgkins} is Program Director at the Center for AI Safety.

\textbf{James Petrie} is Compute Security \& Governance Specialist at the Future of Life Institute

\textbf{Aydin Aysu} is an Associate Professor at North Carolina State University.

\textbf{Kanad Basu} is an Assistant Professor in the Department of Electrical and Computer Engineering at the University of Texas at Dallas.

\textbf{Shivam Bhasin} is an Principal Research Scientist at Nanyang Technological University, Singapore.

\textbf{Vincent Immler} is an Assistant Professor of Electrical and Computer Engineering at Oregon State University.

\textbf{Ankur Srivastava} is a Professor in the Department of Electrical and Computer Engineering at the University of Maryland and director of the Institute for Systems Research.

\textbf{Stjepan Picek} is an Associate Professor at Radboud University, The Netherlands.

\section*{Acknowledgements}

This report is based on a workshop organized in August 2024 by the Center for AI Safety.

We would like to thank workshop participants who were unable to take part in the writing of this report but contributed greatly to the discussion: Anthony Aguirre, Berk Sunar, Caleb Withers, Chau-Wai Wong, Daniel Holcomb, Patrick Schaumont, Sanjit Seshia, Simha Sethumadhavan, Tim Fist and Waleed Khalil.

\end{document}